\documentclass[a4paper,11pt]{article}
\pdfoutput=1
\usepackage{jinstpub}
\usepackage{siunitx}
\usepackage{tikz}
\usetikzlibrary{arrows.meta}
\usepackage[acronym]{glossaries}
\DeclareSIUnit\lightspeed{c}
\DeclareSIUnit\tev{\tera\electronvolt}
\DeclareSIUnit\gev{\giga\electronvolt}
\DeclareSIUnit\mev{\mega\electronvolt}
\DeclareSIUnit\kev{\kilo\electronvolt}
\DeclareSIUnit\year{years}
\DeclareSIUnit\nothing{\relax}

\newcommand*{\pt}{p_\mathrm{T}} 

\title{\boldmath To catch a long-lived particle: hit selection towards a regional hardware track trigger implementation}

\author{M. M\aa rtensson,}
\author{M. Isacson,}
\author{H. Hahne, }
\author{R. Gonzalez Suarez }
\author{and R. Brenner}
\affiliation{Uppsala universitet, L{\"a}gerhyddsv{\"a}gen 1, 752 37 Uppsala, Sweden}

\emailAdd{rebeca.gonzalez.suarez@physics.uu.se}

\abstract{

Conventional searches for new phenomena at collider experiments tend to focus
on prompt particles, produced at the interaction point and decaying rapidly.
New physics models including long-lived particles that travel a substantial
distance in the detectors before decaying provide an interesting alternative,
especially in light of the lack of new phenomena at the current LHC
experiments, and could solve unanswered
questions of the Standard Model. Long-lived particles have characteristic
experimental signatures that, while making them clearly distinct from other
processes, also could make them potentially invisible to current
data-acquisition methods. Specific trigger strategies need to be in place to
target long-lived particles. In this paper, we investigate the use of tracker
information at trigger level to identify displaced signatures. We propose two
methods that can be implemented at hardware-level: one based on the Hough
transform, and another based on pattern matching with patterns trained on
displaced tracks.

}


\arxivnumber{1907.09846}

\newacronym{HL-LHC}{HL-LHC}{High-Luminosity Large Hadron Collider}
\newacronym{LHC}{LHC}{Large Hadron Collider}
\newacronym[longplural={Regions of Interest}]{RoI}{RoI}{Region of Interest}
\newacronym{SM}{SM}{Standard Model}
\newacronym{LLP}{LLP}{long-lived particles }
\newacronym{GPU}{GPU}{Graphics Processing Unit}
\newacronym{FPGA}{FPGA}{Field-Programmable Gate Array}
\newacronym{SSW}{SSW}{Super Strip Width}
\newacronym{HTT}{HTT}{Hardware-based Tracking for the Trigger}

\begin{document}
\maketitle
\flushbottom

\section{Introduction}

The different particles of the Standard Model (SM) display a wide range of
lifetimes, from the very short, like the top quark (\SI{0.5e-24}{\s}), to the
very long, like the proton ($>\SI{2.1e29}{\year}$)~\cite{PhysRevD.98.030001}.
In the same way, particles predicted by theories Beyond the SM (BSM) have
different lifetimes. Particles with lifetimes long enough to give rise to
distinct experimental signatures, such as detectable displaced vertices,
feature in many of those. The unique signatures of such long-lived particles
(LLPs) offer exciting opportunities for discovery of new physics at collider
experiments, but at the same time they require specific detection strategies
and dedicated searches~\cite{Alimena:2019zri} since they can be easily missed.

In this paper, we explore two methods, one based on the Hough transform, and
another on pattern matching, to identify displaced tracks.
These two methods can be implemented in hardware to be used to
trigger on displaced signatures, targeting events containing LLPs for the High
Luminosity upgrade of the LHC (HL-LHC)~\cite{Apollinari:2017cqg}. This study
is performed using a subset of layers of the tracker,
corresponding to what is referred to as ``regional tracking'' in the Hardware
Tracking for the Trigger (HTT) for the ATLAS upgrade~\cite{ATLAS-Phase2}.

\section{Motivation}

Most direct searches for new physics at the LHC experiments focus on particles
with short lifetimes. The LHC experiments were optimized to detect decays of
such particles, like the Higgs boson, that decays almost immediately after
being produced, with a lifetime on the order of \SI{e-22}{\s}. However, in the
search of BSM effects at the LHC and the HL-LHC, LLPs provide a promising
alternative~\cite{Lee:2018pag}.

There is a strong theoretical motivation for LLP searches. Models including
LLPs could provide answers to central questions still unanswered by the SM,
such as naturalness~\cite{Cai:2008au}, dark
matter~\cite{ArkaniHamed:2008qn,Pospelov:2008jd},
baryogenesis~\cite{Ipek:2016bpf}, and neutrino masses~\cite{Antusch:2016vyf},
among others. This variety of LLP models cover a broad range of lifetimes,
production mechanisms, and decay products, making the search for LLPs a very
rich experimental area. Although LLPs are vastly unexplored they have been
studied in previous colliders, and many theoretical models are under ongoing
scrutiny or have been searched for at the LHC. Nevertheless, many others, such
as dark showers~\cite{Cohen:2017pzm} remain to be explored.

Unique signatures provide unique challenges, and there are different ways to
improve our ability to record and analyze LLPs. Dedicated detectors, far from
the interaction point, such as MATHUSLA~\cite{Lubatti:2019vkf}, FASER~\cite{Ariga:2018uku}, or
CODEX-b~\cite{Gligorov:2017nwh}, are one option. Another option is to
optimize the trigger and reconstruction capabilities of the existing
all-purpose detectors. Along those lines, ATLAS is currently investigating the reconstruction of 
LPPs in the trigger for Run 3~\cite{Hooberman:2677874} and the potential of CMS to identify this kind of signatures at trigger-level at the HL-LHC has been highlighted in Reference~\cite{Gershtein:2017tsv}.

The HL-LHC will provide challenging experimental conditions, with higher
pile-up\footnote{Pile-up is the mean number of simultaneous proton-proton
interactions. A pile-up of approximately 200 is expected at \gls*{HL-LHC}.} and
occupancy, and much higher readout rates. In particular, the trigger and tracking
systems of the HL-LHC experiments will be crucial for LLP searches. Dedicated
LLP triggers are justified, since important signatures could be missed without
them, such as neutral LLP decays within the tracker, or LLP with
low-$p_\text{T}$ objects in general. Ongoing LLP searches in ATLAS and CMS
are already limited by the trigger systems.

In this paper we focus on one of the most pressing needs in this area, the use
of tracker information at trigger-level to identify LLPs. Track information
was not available at trigger-level in Run-1 and Run-2 in ATLAS or CMS, but will
start being used for Run-3 and beyond. Assuming a generic tracker composed of five
pixel layers up to \SI{300}{\mm} and five strip layers up to \SI{1000}{\mm}, we
can identify three kinds of signatures that we can target:
\begin{itemize}
    \item \textbf{Detector-stable LLPs}. These are characterized by tracks that show
        anomalous ionization, i.e. a different energy loss pattern in the
        tracker than common SM particles. This includes the so-called heavy,
        stable charged particles (HSCPs), and magnetic monopoles. These kinds of
        LLPs have long decay lengths ($c\tau > \SI{1000}{\mm}$).
    \item \textbf{Disappearing tracks}. Occur when a charged LLP decays inside
        the tracker (decay length $c\tau \approx \SI{100}{\mm}$) into a neutral
        state and a soft SM charged particle, making it look like the original
        track has ``disappeared''.
    \item \textbf{Displaced vertex/tracks}. This kind of signature occurs when
        a track displays a large transverse vertex displacement, incompatible with
        being produced at the interaction point. If more than one such track
        exists and they share a common point of origin, this conforms to a
        displaced secondary vertex. This kind of signatures have decay lengths of $c\tau \approx $~\SIrange{100}{1000}{\mm}, and are the ones we are
        targeting in this paper. The exploration of the other two should be
        pursued also later on.
\end{itemize}

\begin{figure}
  \centering
  \includegraphics[width=0.6\columnwidth]{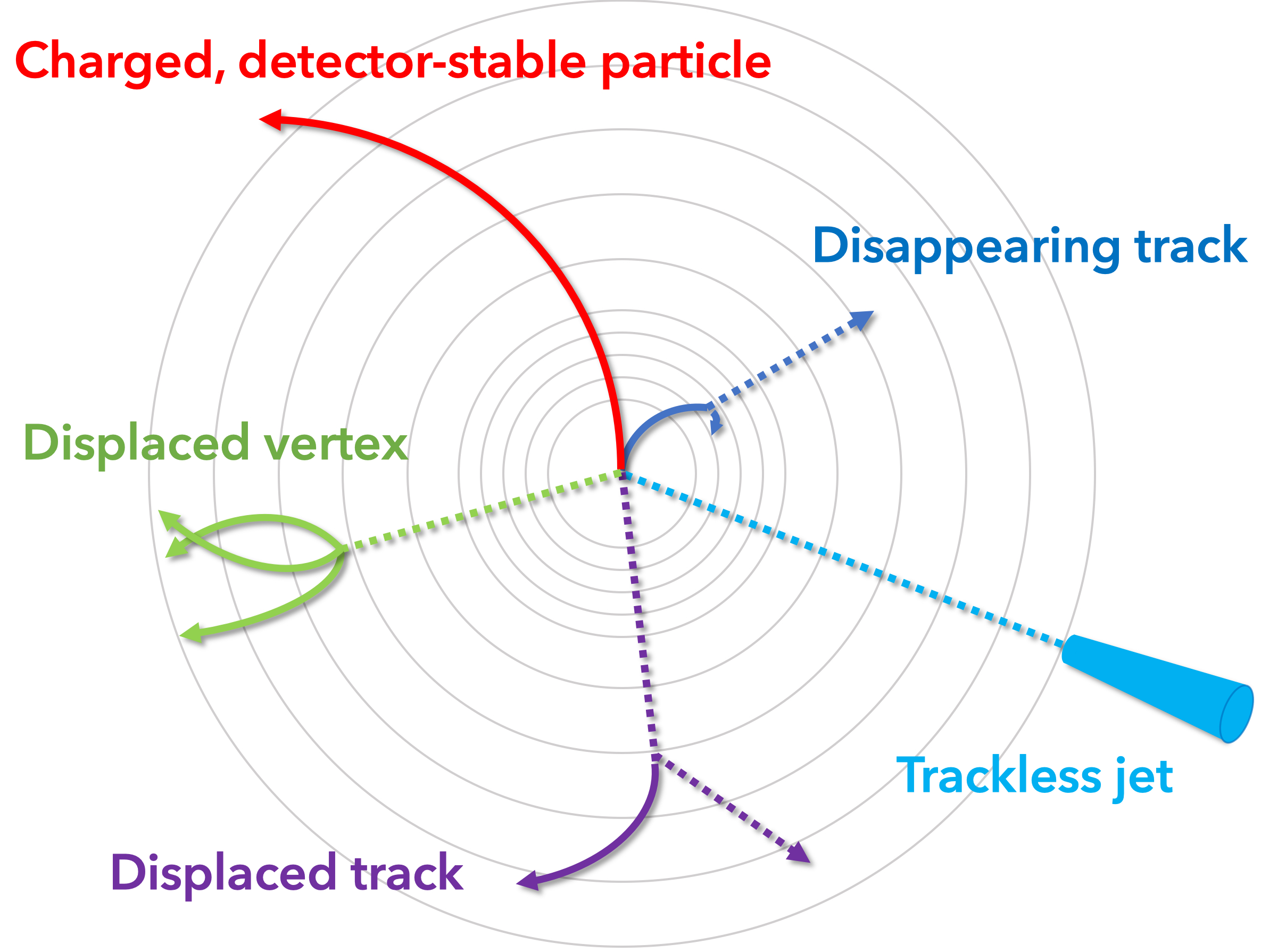}
  \caption{Different processes involving LLP signatures in a generic tracker, represented by 5 pixel layers and 5 strip layers.}
  \label{fig:cartoon}
\end{figure}

Another LLP signature which can profit from tracking information at trigger
level are neutral LLPs decaying in the calorimeters, leaving trackless jets.
Figure~\ref{fig:cartoon} shows a schematic sketch of these processes.

For the decay lengths of up to $\SI{350}{\mm}$ (within the strip detector), which are targeted in this paper, there are a number of scenarios that could be explored: split-SUSY, R-Parity
Violating (RPV) SUSY, Gauge Mediated Supersymmetry Breaking (GMSB), or Hidden
Valley to name a few. However, we will not focus on a particular physics model
at this point. Instead, we will focus on methods that could be used in
hardware-based processing to trigger displaced signatures, no matter their
underlying production mechanism. This could be used to define dedicated
displaced triggers for specific signatures, but also, maybe more importantly,
it could be used in combination with other triggers to reduce thresholds,
something that can be critical for the detection of LLPs at the HL-LHC.

\section{Simulation} \label{sec:simulation}

\begin{figure}[htbp]
    \centering
    \includegraphics[width=0.7\columnwidth]{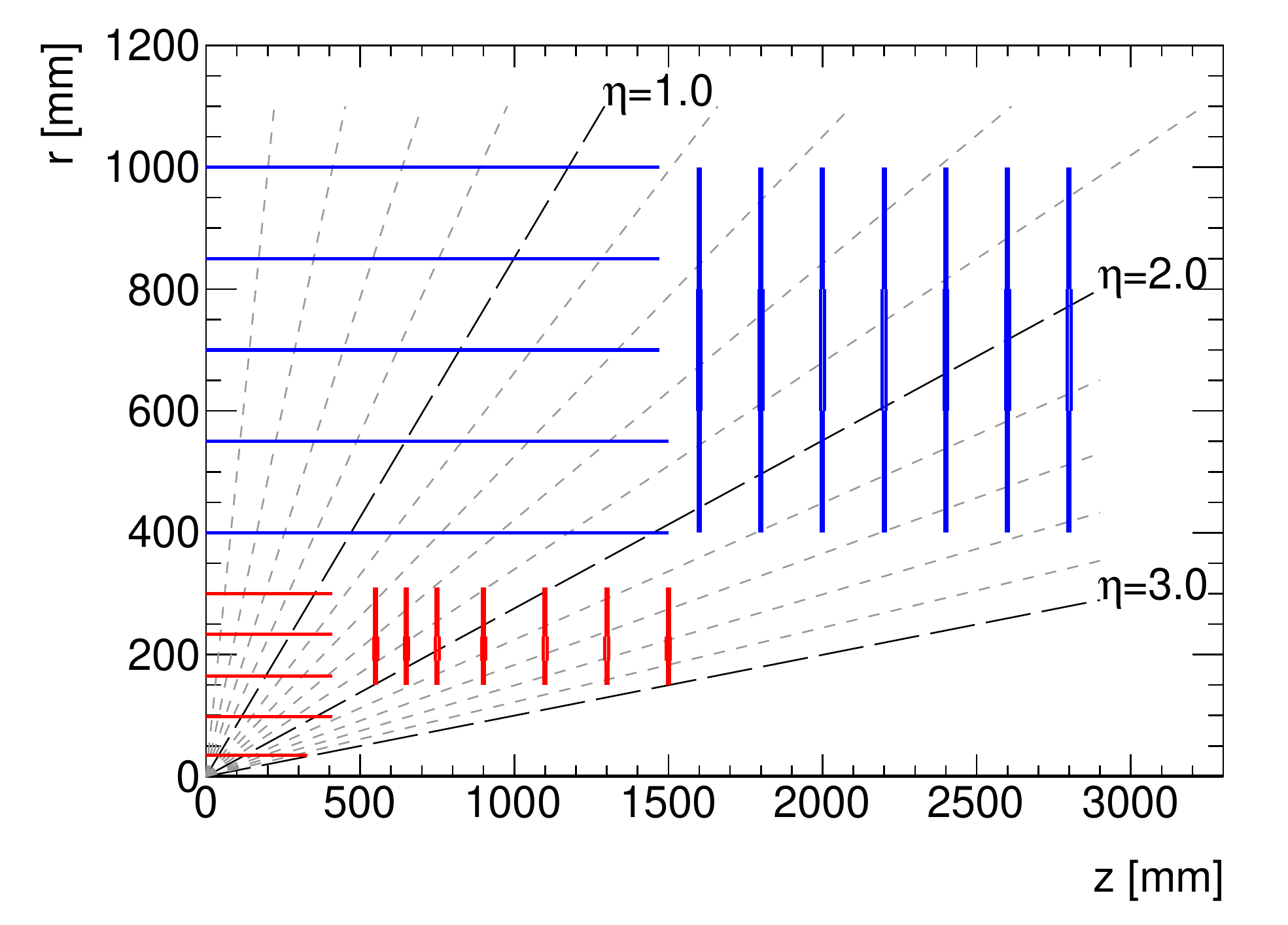}
    \caption{Tracker geometry. The detector consists of 5 (7) pixel layers, shown in red, and 5 (7) strip double layers, shown in blue, in the barrel (end-cap).~\cite{Gradin_2018}}\label{fig:geometry}
  \end{figure}

A generic tracker similar to that used in ATLAS and CMS is modelled using the
\textsc{Geant4}~\cite{Agostinelli:2002hh} toolkit. The simulation uses a right-handed coordinate system with the $z$-axis pointing along the beam-line. The $x$- and $y$-axes make up transverse plane. The geometry is the same as
that used in Reference~\cite{Gradin_2018} and depicted in
Figure~\ref{fig:geometry}. The barrel consists of five layers of pixel
sensors, evenly spaced between \SI{35}{\milli\meter} and
\SI{300}{\milli\meter}, and five double layers of strip sensors, evenly spaced
between \SI{400}{\milli\meter} and \SI{1000}{\milli\meter}.  The end-caps
consist of seven layers of pixel and seven double layers of strip sensors.
Both the pixel and silicon sensors are modelled as \SI{320}{\micro\meter} thick
silicon wafers. The active area of the pixel sensors is
\SI{40x40}{\milli\meter}, except in the innermost barrel and end-cap layers
where the active width is \SI{20}{\milli\meter} and the outermost end-cap layer
where the active width is \SI{80}{\milli\meter}.  The active length of the
strip sensors is \SI{24}{\milli\meter} in the two innermost barrel double
layers, and \SI{48}{\milli\meter} in the three outermost barrel double layers.
The active width of the barrel modules is \SI{100}{\milli\meter}. In the
end-cap layers the active width is \SI{100}{\milli\meter} except in the two
outermost double layers where the active width is \SI{200}{\milli\meter}. The
active length of all strip sensors in the end-caps is \SI{100}{\milli\meter}.
The two single layers in each strip double layer are spaced
\SI{5}{\milli\meter} apart and rotated by \SI{20}{\milli\radian} about their
geometrical center.

The continuous local hit coordinates $(x,y)$ are recorded modulo the readout
granularity of the respective sensor~\cite{Gradin_2018}. The pixel granularity is
\SI{25x250}{\micro\meter} and the granularity of the strip sensors is
$\SI{80}{\micro\meter}$ in the $x$-direction.

Tracks are described by helix parameters given by the position, momentum, and trajectory of the track at its closes approach to the $z$-axis: the transverse momentum ($p_\mathrm{T}$), the azimuthal angle ($\phi_0$), the pseudorapidity\footnote{The pseudorapidity is defined as $\eta = -\ln{\tan{\left(\theta/2\right)}}$, where $\theta$ is the polar angle in the right-handed spherical coordinates.} ($\eta_0$), the transverse impact parameter ($d_0$), and the longitudinal impact parameter ($z_0$) of the track. The coordinates of the production vertex of a track is given by ($x_\mathrm{V}, y_\mathrm{V}, z_\mathrm{V}$), or the radial coordinate $d_\mathrm{V}=\sqrt{x_\mathrm{V}^2 + y_\mathrm{V}^2}$ and the azimuthal coordinate $\phi_\mathrm{V}$. The directions of a track at the production vertex are denoted $\phi$ and $\eta$.

Muon tracks from displaced vertices are generated in a model-independent way
using a particle gun implemented with
\textsc{Pythia}~8.2~\cite{Sjostrand:2014zea}. The position of the track vertex
is determined by drawing the longitudinal, transverse, and azimuthal
coordinates from flat distributions between \SIrange{-150}{270}{\milli\meter},
\SIrange{0}{350}{\milli\meter}, and \SI{+-\pi}{\radian} respectively.

\begin{figure}
    \centering
    \begin{minipage}{0.4\linewidth}
        \includegraphics[width=\columnwidth]{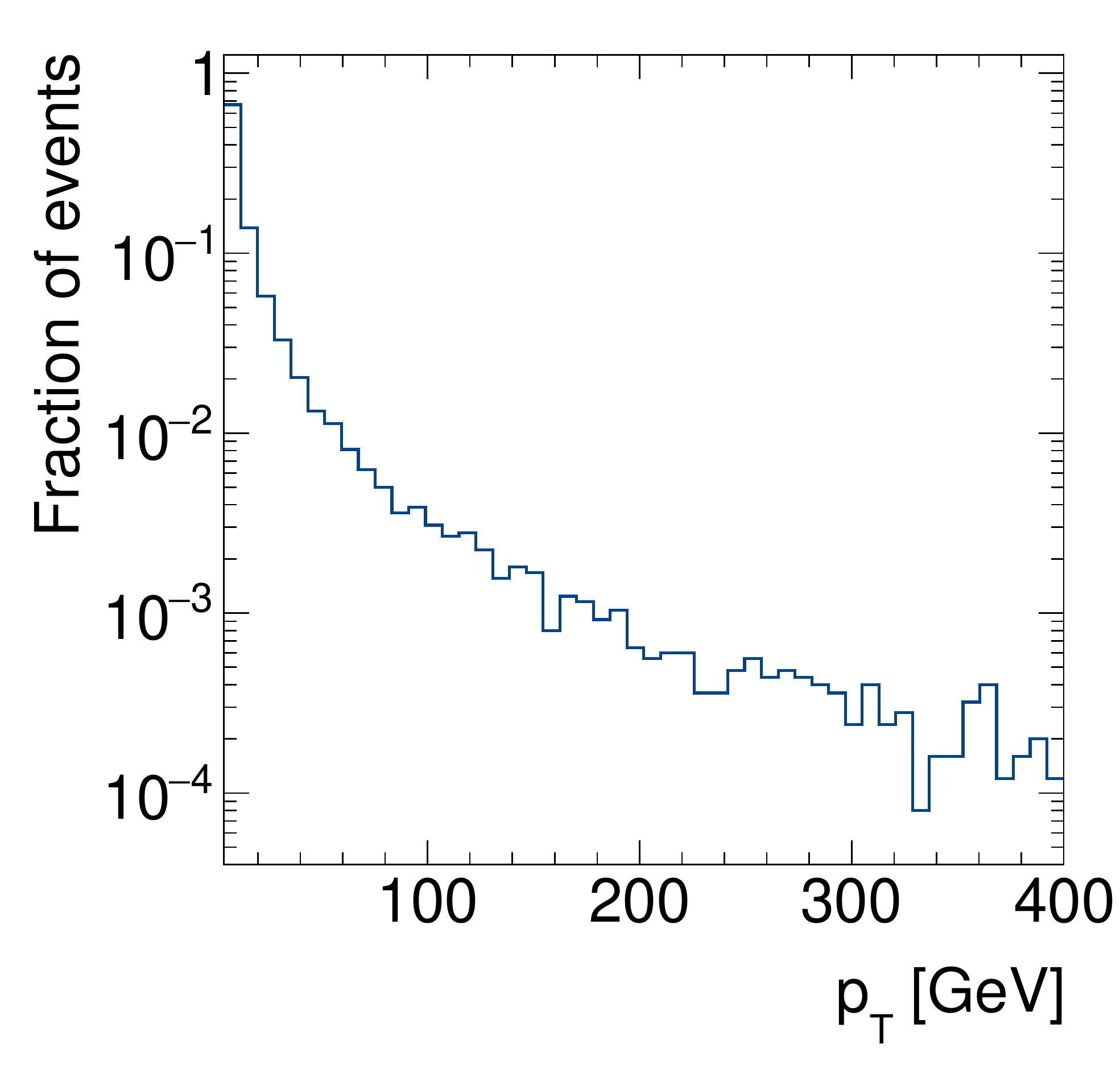}
    \end{minipage}
    \hspace{1em}
    \begin{minipage}{0.4\linewidth}
        \includegraphics[width=\columnwidth]{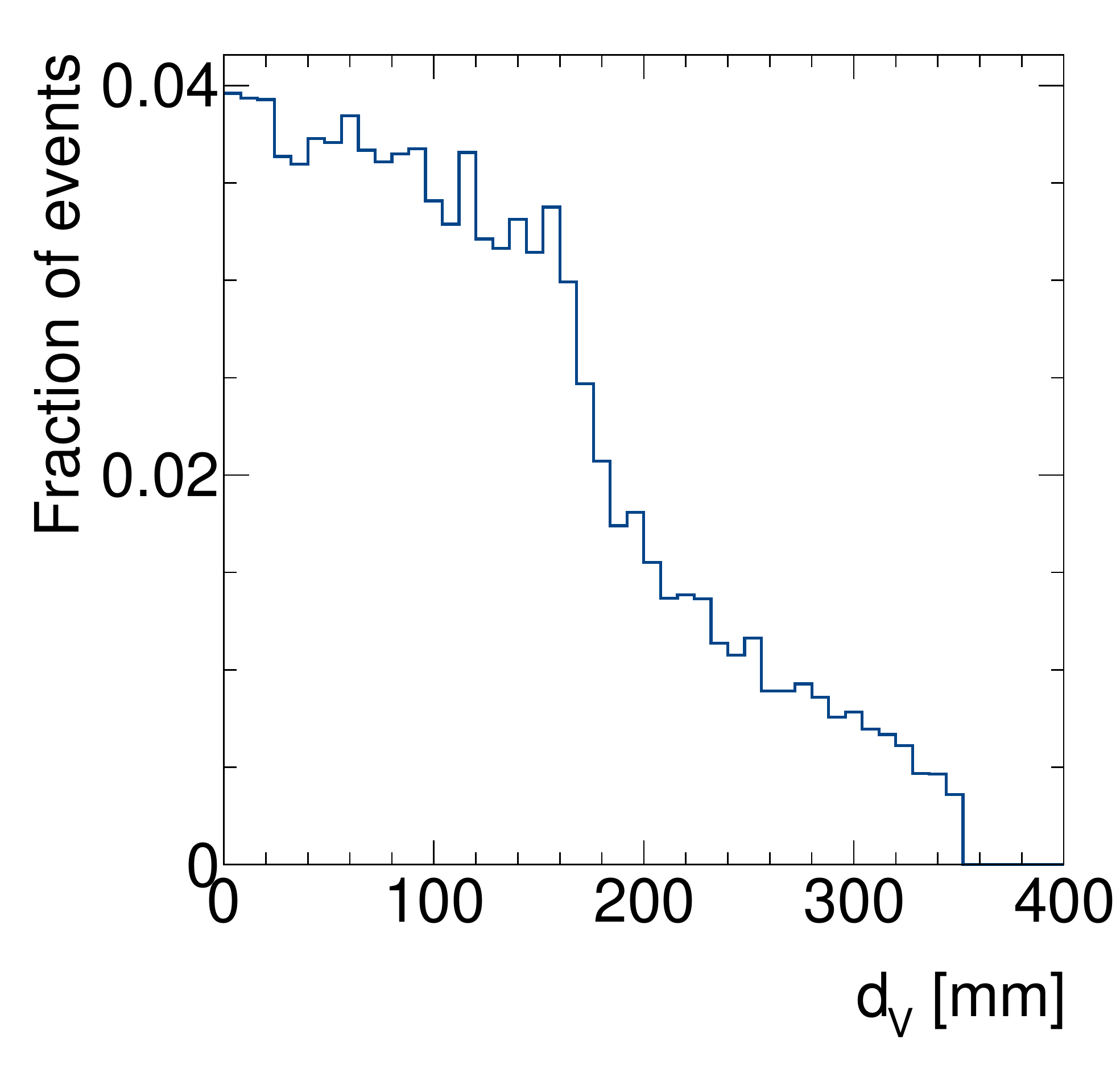}
    \end{minipage}
    \caption{Sample distributions of the transverse momentum (Left) and transverse vertex displacement (Right)
            of the generated muons.}
    \label{fig:pdf_pt_d0}
\end{figure}

\begin{figure}
    \centering
    \begin{minipage}{0.4\linewidth}
        \includegraphics[width=\columnwidth]{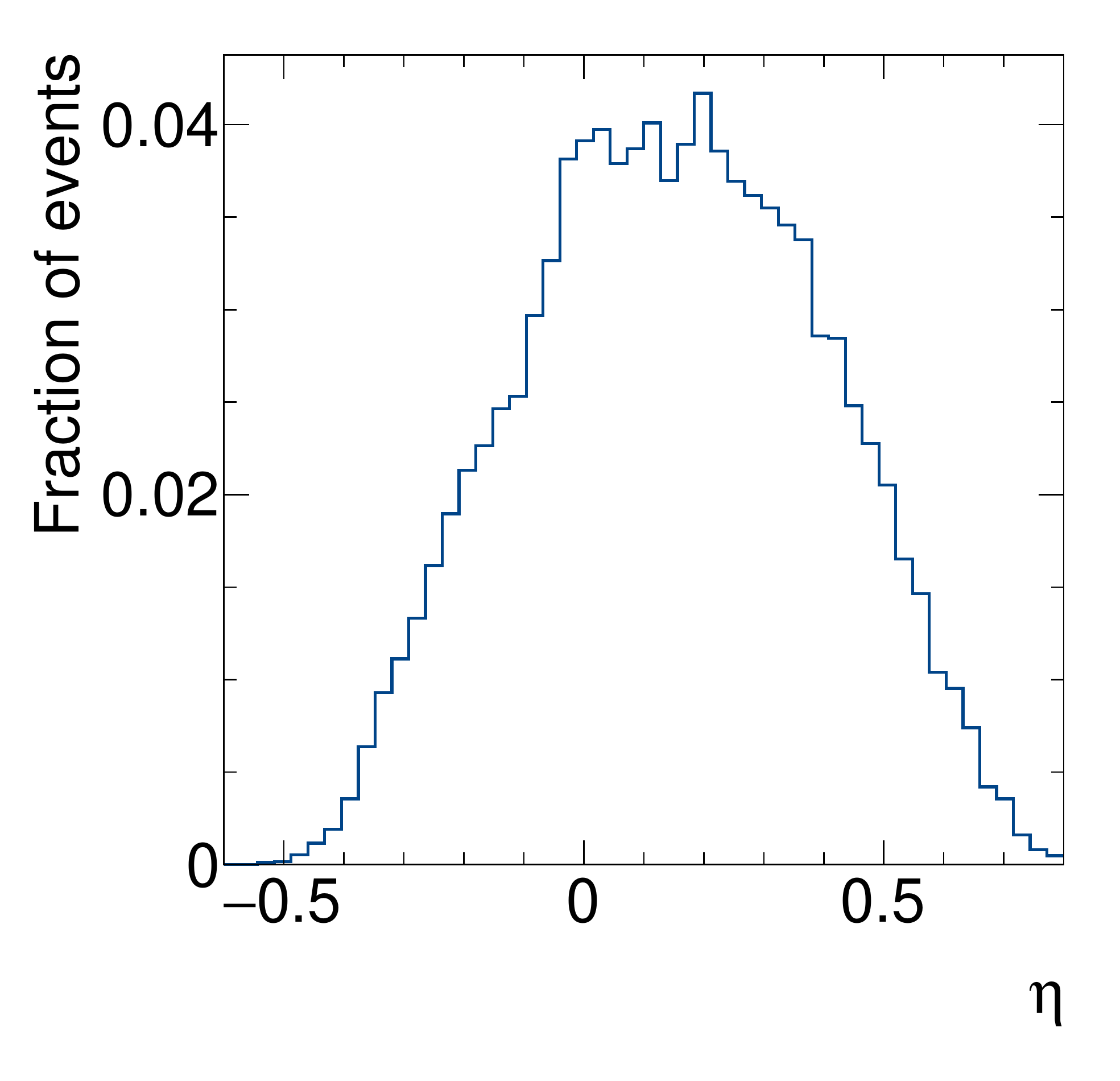}
    \end{minipage}
    \hspace{1em}
    \begin{minipage}{0.4\linewidth}
        \includegraphics[width=\columnwidth]{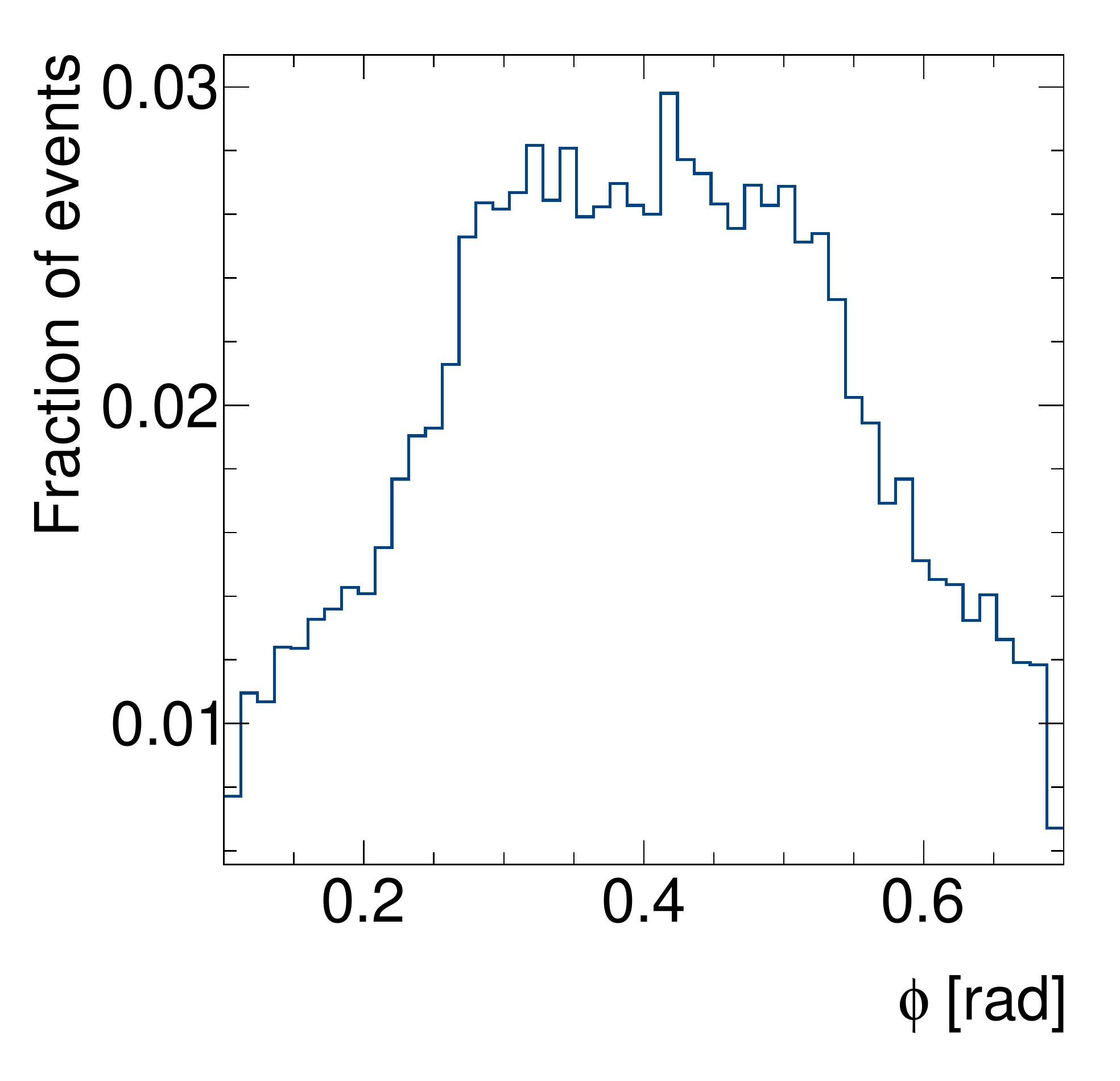}
    \end{minipage}
    \caption{Sample distributions of the pseudorapidity (Left) and azimuthal angle (Right)
            of the generated muons.}
    \label{fig:pdf_eta0_phi0}
\end{figure}

\begin{figure}
    \centering
    \begin{minipage}{0.48\linewidth}
        \includegraphics[width=\columnwidth]{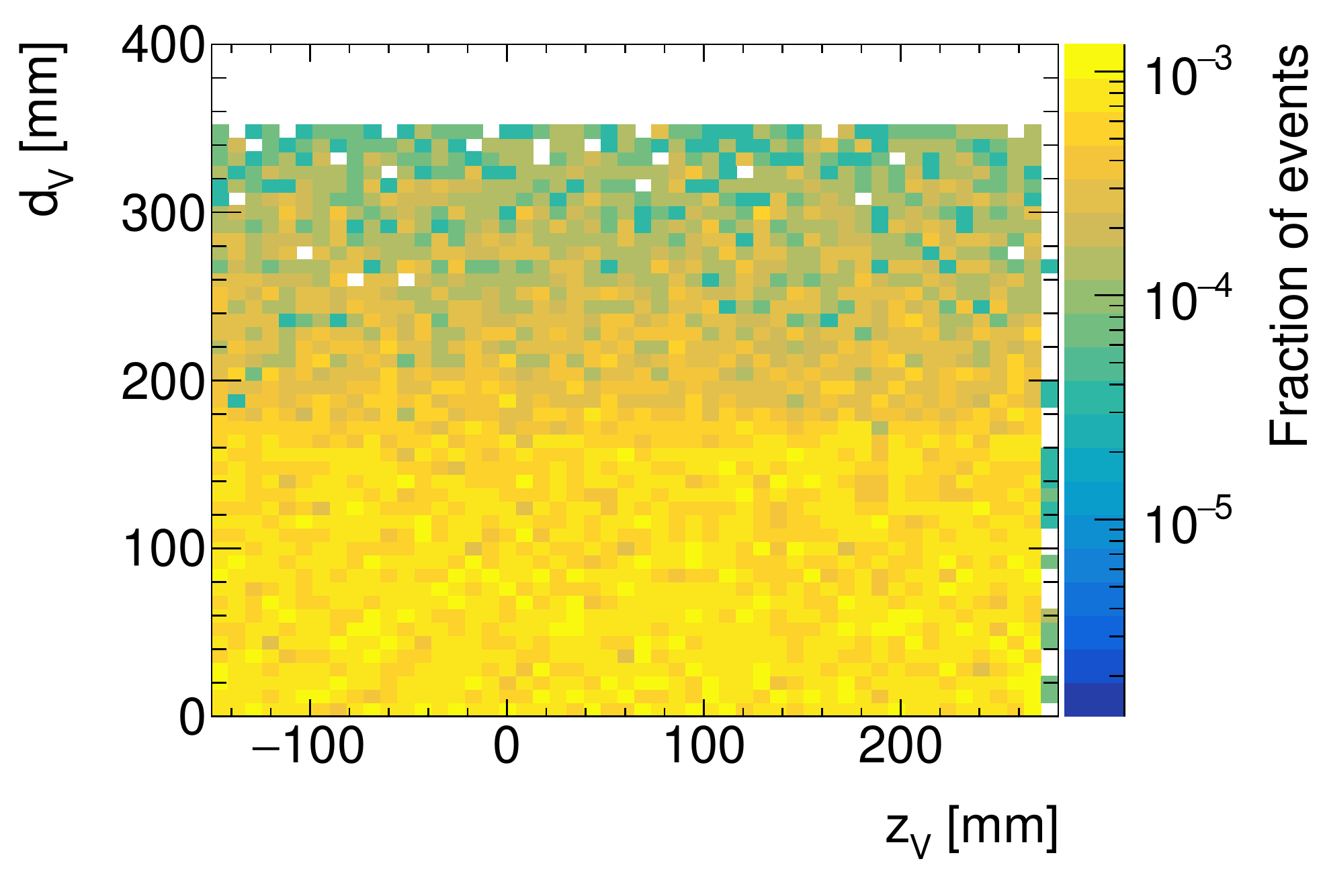}
    \end{minipage}
    \hspace{1em}
    \begin{minipage}{0.48\linewidth}
        \includegraphics[width=\columnwidth]{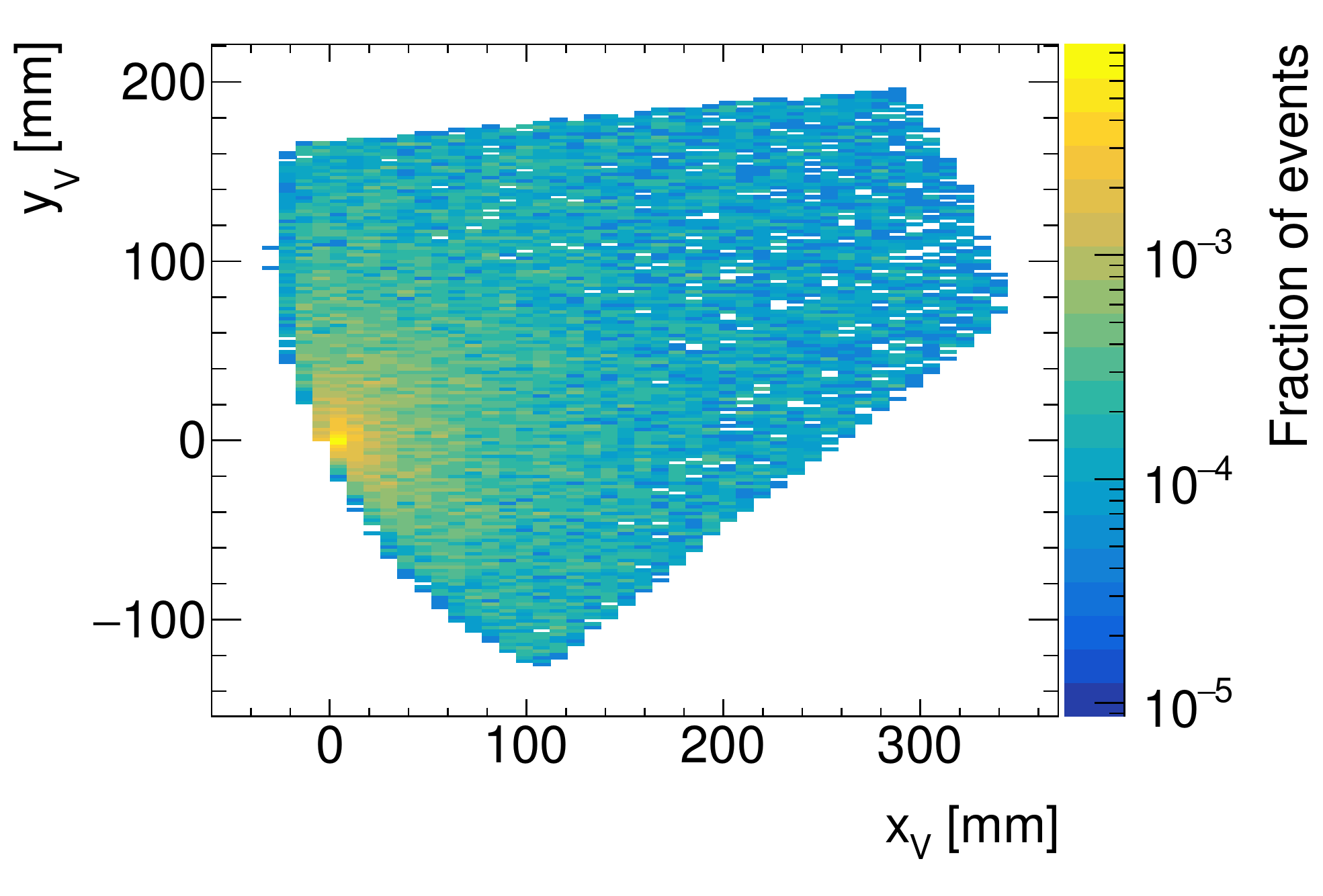}
    \end{minipage}
    \caption{Sample distributions of the vertex positions in the
            longitudinal plane (Left) and transverse plane (Right).}
    \label{fig:pdf_vtx_position}
\end{figure}

 The tracks are generated such that they pass through all strip layers in the
 \gls*{RoI} defined by the geometric volume covered by prompt tracks with $0.1
 < \eta_0 < 0.3$, $0.3 < \phi_0 < 0.5$, $|z_0| < \SI{150}{\mm}$, and
 $p_\mathrm{T}>\SI{4}{\GeV}$.  This is the reason for the asymmetric bounds on
 the $z_\mathrm{V}$ distribution. The upper bound needs to be extended to allow for full
 coverage for the strip modules in the \gls*{RoI}.  Tracks are rejected if
 $|\phi - \phi_\mathrm{V}| > \frac{\pi}{2}$ to bias against tracks going
 backwards toward the primary vertex. The transverse momentum $\pt$ is drawn
 between \SIrange{4}{400}{\gev} from a flat distribution in $1/\pt$. In total
 \SI{60}{\mega\nothing} tracks are generated split equally between $\mu^+$ and
 $\mu^-$. Setting the bounds on the muon direction dynamically and rejecting
 tracks pointing toward the primary vertex skews the resulting distributions of
 the track direction and vertex position, shown in Figure~\ref{fig:pdf_pt_d0} (Right),
 Figure~\ref{fig:pdf_eta0_phi0}, and Figure~\ref{fig:pdf_vtx_position}.
 However, the bias towards lower $d_\mathrm{V}$ should not pose a serious problem. The
 distribution is reasonably flat up to \SI{\sim150}{\milli\meter}, and the
 exponentially distributed decay length of an LLP will be similarly biased
 towards lower values.

This study uses hits in the outer layers of the tracker to get a maximum number of
hits from displaced tracks. Eight layers of the strip detector are used: the six
outer layers, i.e.\ the three outer double layers, and the outer layer of the
two innermost double layers. The remaining tracker layers are not used but the
material will affect the response in the outer layers. As previously stated,
this corresponds to the regional tracking scenario of the HTT, used to perform
fast initial trigger decisions after Level-0~\cite{ATLAS-Phase2}. 

In the regional HTT the readout of the tracker is seeded by the calorimeters or muon detectors. Since at most 10\% of the tracker volume will be read out at any time, the regional tracking can be run at a higher rate than the full readout of the tracker. This allows for lower thresholds on particles used in a regional trigger than for a trigger based on full readout.

\section{Filtering of tracker hits}
Here the two hardware based hit collection methods studied in this paper are
described. Section~\ref{sec:hough} describes the Hough transform method, and
Section~\ref{sec:pattern} describes the pattern matching method.

\subsection{Hough transform method}\label{sec:hough}

The Hough transform was invented in 1959 to analyze photographic plates of
bubble chambers~\cite{hough}. Since then, the method has found use in various
computer vision applications. With recent developments in GPU and FPGA
technology, the interest in the Hough transform for tracking applications in
high energy physics has been renewed. The Hough transform can be constructed
for any curve that can be described with a few parameters. For each spatial
point in image-like data, it calculates one of the unknown parameters from the
parameterization and the point coordinates while sweeping the other parameters.
It then casts votes in a histogram-like object called the \emph{accumulator}.
Points in the accumulator with a lot of votes are candidates realizations of
the searched-for curve in the data.

Charged particles trace out helices as they propagate through a uniform
magnetic field. Viewing the track along the magnetic field lines, in what is
called the transverse plane of the detector, they appear as circular arcs. The
radius of curvature is proportional to the transverse momentum of the particle.
A track starting at the origin and passing through a point $(r,\phi)$ can be described by the
transverse momentum $p_\textrm{T}$ and the azimuthal angle $\phi_0$ according
to

\begin{equation}
  \label{eq:hough}
  A\frac{qB}{p_\text{T}} = \frac{\sin{(\phi_0-\phi)}}{r}\,,
\end{equation}

\noindent where $q$ is the elementary charge, $B$ is the electromagnetic field
strength, and $A$ is a unit conversion constant with value $A \approx
\SI{1.5e-4}{\GeV\per\lightspeed\per\mm\per\tesla}$. This equation is used as
the basis of the Hough transform studied in this paper. It is straightforward to loosen the
vertex constraint imposed by Equation~\ref{eq:hough}. The downside
to doing so is that the Hough transform needs to be applied to pairs of hits,
which increases the number of computations needed dramatically. This paper aims
to investigate the performance of the Hough transform when keeping it as simple
and computationally inexpensive as possible. Therefore, the vertex constrained
variant is the one studied.

The accumulator is constructed similarly to a 2-dimensional histogram with
$\phi_0$ on one axis and $AqB/p_\text{T}$ on the other. The bins in the
accumulator keeps track of which layers have hits that are consistent with a
track parameterized by the $\phi_0$ and $AqB/p_\text{T}$ of that bin.
Each bin also keep track of which hits have been used to fill it. The goal here
is not to use the track parameters $p_\mathrm{T}$ and $\phi_0$ from the Hough
transform as is, but rather to select hits to send to a subsequent, dedicated
track fitter, for instance, a linearized track-fit using precomputed constants, such as the one planned for the ATLAS Phase-II upgrade~\cite{ATLAS-Phase2}.
Hits are selected from a single bin in the accumulator by applying a threshold on the number of hits. The requirement is that the bin should have at least six hits and, in addition, that the two nearest neighbouring bins along constant $AqB/p_\text{T}$ should have at least five hits, and that the next-to-nearest neighboring bins along constant $AqB/p_\text{T}$ should have more than four hits.

The Hough transform is applied in small \glspl*{RoI} of $0.2\times 0.2$ in
$\Delta \eta \times \Delta \phi_0$. However, the regions are \SI{300}{\mm} wide
in the longitudinal impact parameter $z_0$. To improve the rejection of
unwanted soft-QCD background, the \glspl*{RoI} are split into several slices in
$z_0$. This and more details of the Hough transform implemented here are
described in Reference~\cite{Gradin_2018}. Here, we apply the Hough transform
to tracks with displaced vertices.

The results are presented in terms of the efficiency to detect tracks from
signal and the power of rejecting the unwanted soft QCD background. The
efficiency is studied using muon tracks generated in the way described in
Section~\ref{sec:simulation}. A track is considered found if at least six hits
originating from the track pass the Hough transform selection. The hits are required to
be in unique detector layers and to belong to the same bin in the accumulator.
The rejection is studied in terms of the number of possible combinations of
hits that survive the Hough transform. It is calculated as

\begin{equation}
  \label{eq:cluster_combinations}
  \sum\limits_{b} \left( \prod\limits_{l} n_{b,l} \right)\mbox{},
\end{equation}

\noindent where $n_{b,l}$ is the of the number of hits in layer $l$ in accumulator bin number $b$. 

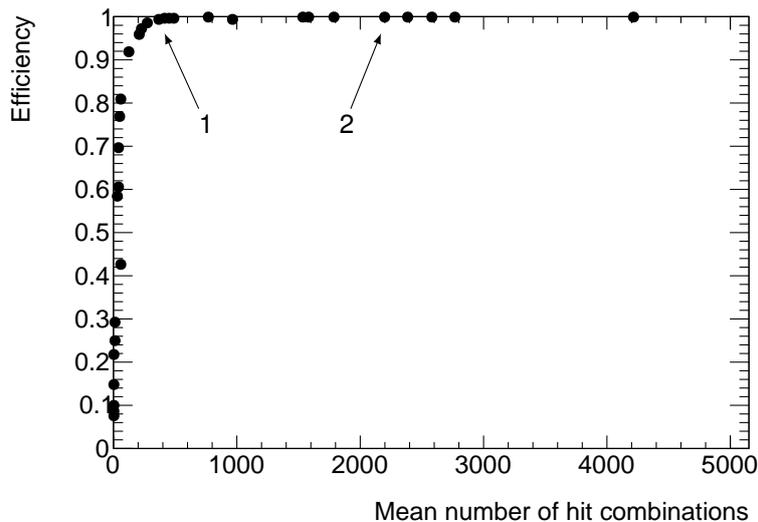
\begin{figure}
  \centering
  \begin{tikzpicture}[font=\sffamily]
    \node[anchor=south west] (pic) at (0,0) {  \includegraphics[width=0.7\columnwidth]{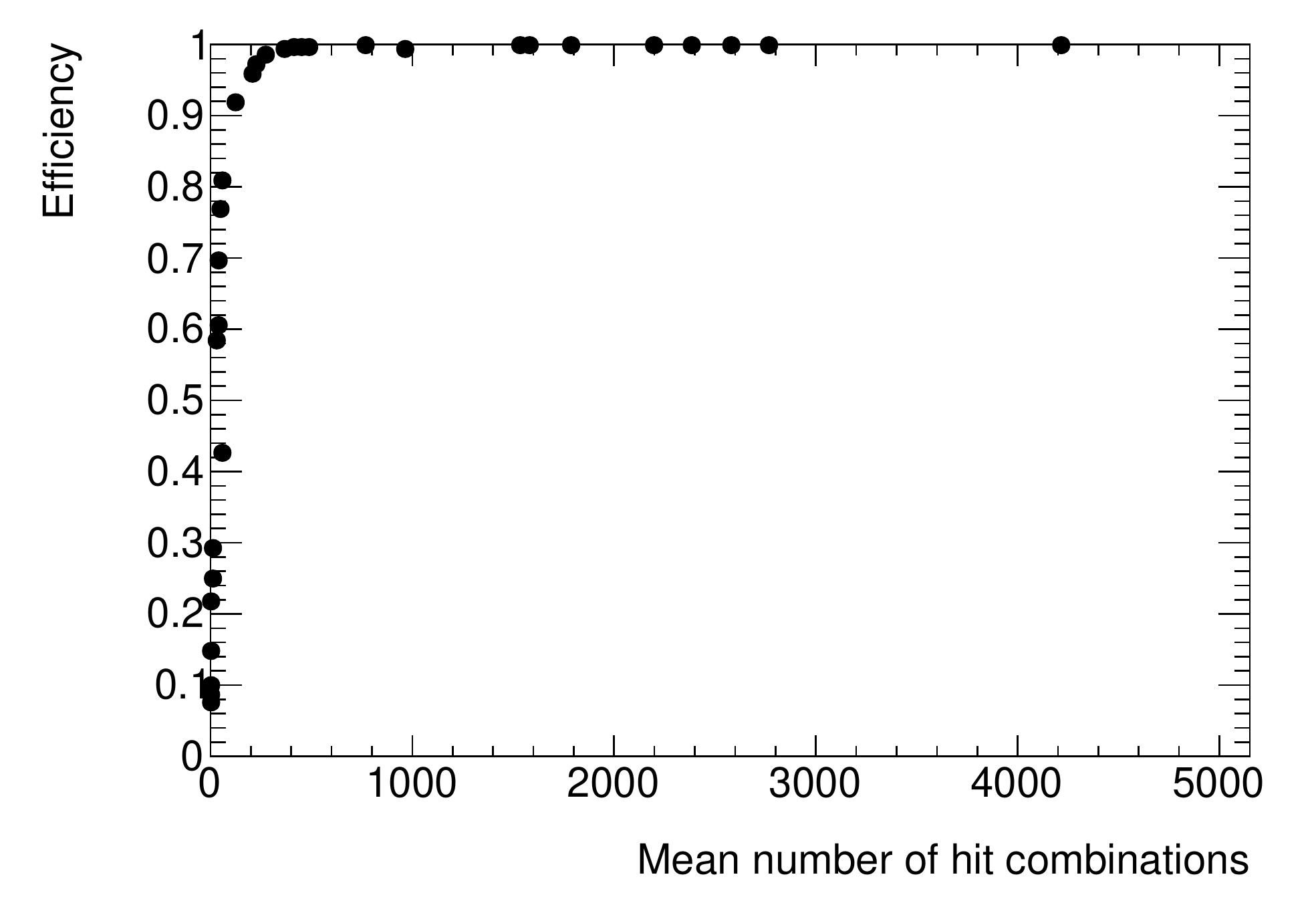}};
    \draw [-latex] (2.95,5.8) -- (2.50,6.8);
    \draw [-latex] (4.95,5.8) -- (5.35,6.8);
    \node (wp1) at (3.05,5.6) {1};
    \node (wp2) at (4.9,5.6) {2};
  \end{tikzpicture}
  \caption{Efficiency of finding hits from simulated tracks of muons embedded in minimum bias with pile-up 200 plotted against the number of hit combinations. Each point is a separate configuration of the Hough transform. The working points annotated 1 and 2 are used in this paper.}
  \label{fig:hough_eff_vs_hit_comb}
\end{figure}

The Hough transform is tuned to provide at least \SI{99}{\%} efficiency of finding tracks from prompt muons ($d_0<\SI{2}{\mm}$) while minimising the number of hit combinations in minimum bias with pile-up 200. This is done by varying the number of bins in $AqB/p_\text{T}$, the number of bins in $\phi_0$, and the number of $z_0$-slices. The muon and minimum bias simulation used for this are the same ones described in Reference~\cite{Gradin_2018}. Figure \ref{fig:hough_eff_vs_hit_comb} shows the efficiency versus number of hit combinations for muon tracks embedded in minimum bias with pile-up 200 in the $0.1 < |\eta| < 0.3$ \gls*{RoI}. Each point is a separate configuration of the Hough transform. Point 1 has just above \SI{99}{\%} efficiency and is selected as the nominal configuration; it has 95 bins in $AqB/p_\text{T}$, 63 bins in $\phi_0$, and 6 splits in $z_0$. Another configuration, point 2 in figure \ref{fig:hough_eff_vs_hit_comb}, is selected as an alternative configuration with higher efficiency. For point 2, the number of bins in $AqB/p_\text{T}$ is 63, otherwise the configuration is the same as point 1.

\subsection{Pattern matching method}\label{sec:pattern}

Hit filtering using pattern matching is a method that can be easily implemented
in hardware-based processing using content addressable memory (CAM)~\cite{AM}.
The method was successfully demonstrated in the fast online tracker for the CDF
silicon vertex trigger~\cite{SVT} at Fermilab. The method is currently being
implemented in the Fast TracKer (FTK)~\cite{FTK} system for the ATLAS
experiment at CERN and is the baseline method to be used for the high
luminosity upgrade of the ATLAS tracker~\cite{ATLAS-Phase2}.

A pattern is a set of hits. In order to efficiently use memory resources, the
hits in a pattern have lower resolution than the tracker clusters of hits. This
is done by grouping a number of tracker strips or pixels into super strips. A
group of 16 tracker strips has a Super Strip Width (SSW) of 16. A typical
signal from a high-$p_\mathrm{T}$ particle generates single hits in different
detector layers. By simulating a large number of single particles in a
$p_\mathrm{T}$ range of interest, a bank of signal patterns can be generated.
The maximum number of patterns that can be stored depend on the size of the
pattern bank. The number of patterns required to reach high track efficiency depends
 on the granularity and $p_\mathrm{T}$ range of the patterns.
With a higher granularity (lower SSW) or larger $p_\mathrm{T}$ range, more
patterns are required to reach high efficiency. If the patterns are too coarse
however, the method looses its power to reduce background. The optimal is hence
to have a pattern granularity which gives high efficiency and few fakes with
given resources to store patterns.

Detailed description of the method used in this study is described in
Reference~\cite{Gradin_2018}.  In the study presented in this paper we investigate
the effect of extending a pattern bank originally trained for prompt muons
produced in the central interaction region with \SI{10}{\%} and \SI{20}{\%}
extra patterns that have been trained with muons originating from displaced
vertices inside the tracker. We restrict the study to a single region of the
tracker and use only its outermost layers. In the same way as for the previous
method, this scenario corresponds to regional tracking in the
HTT~\cite{ATLAS-Phase2}.

\section{Results}

\begin{figure}
  \centering
  \begin{minipage}{0.4\linewidth}
  \includegraphics[width=\columnwidth]{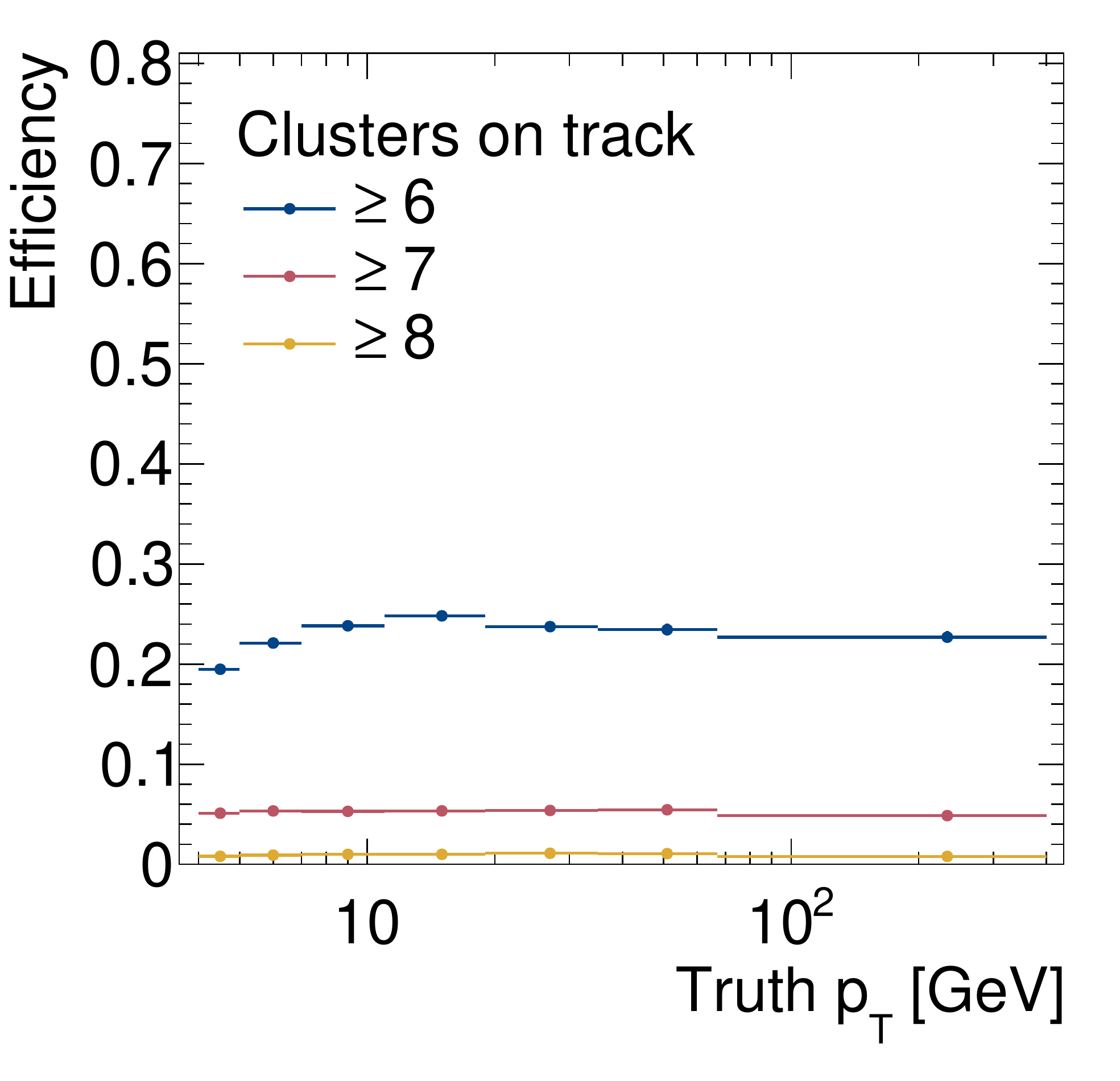}
  \end{minipage}
  \hspace{1em}
  \begin{minipage}{0.4\linewidth}
  \includegraphics[width=\columnwidth]{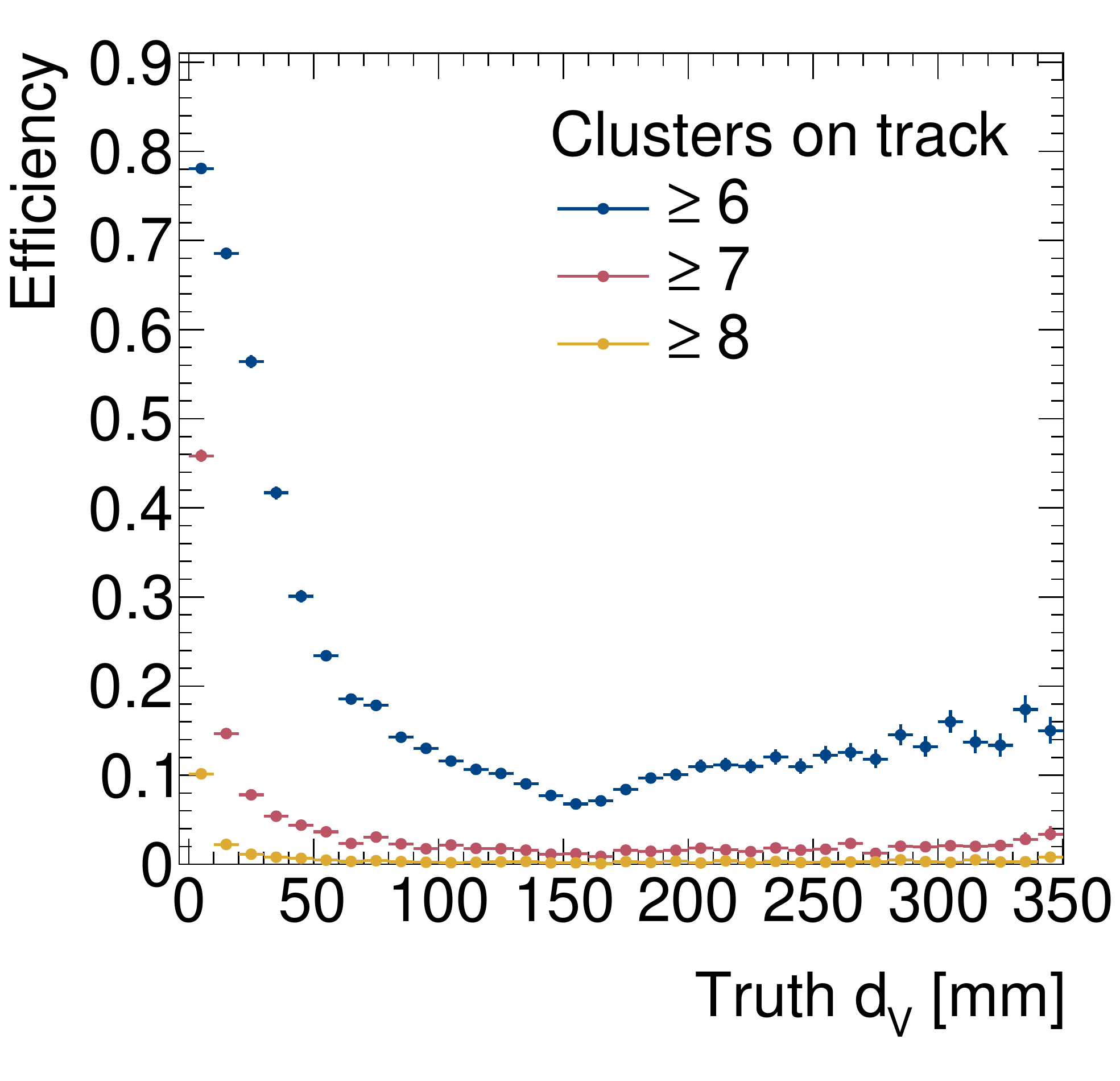}
  \end{minipage}
  \caption{Efficiency of finding at least 6, 7, or 8 hits in unique layers out of 8, as a function of $p_\text{T}$ (Left) and transverse vertex displacement (Right), for configuration 1, Hough transform method.}
  \label{fig:eff_95}
\end{figure}

\begin{figure}
  \centering
  \begin{minipage}{0.4\linewidth}
  \includegraphics[width=\columnwidth]{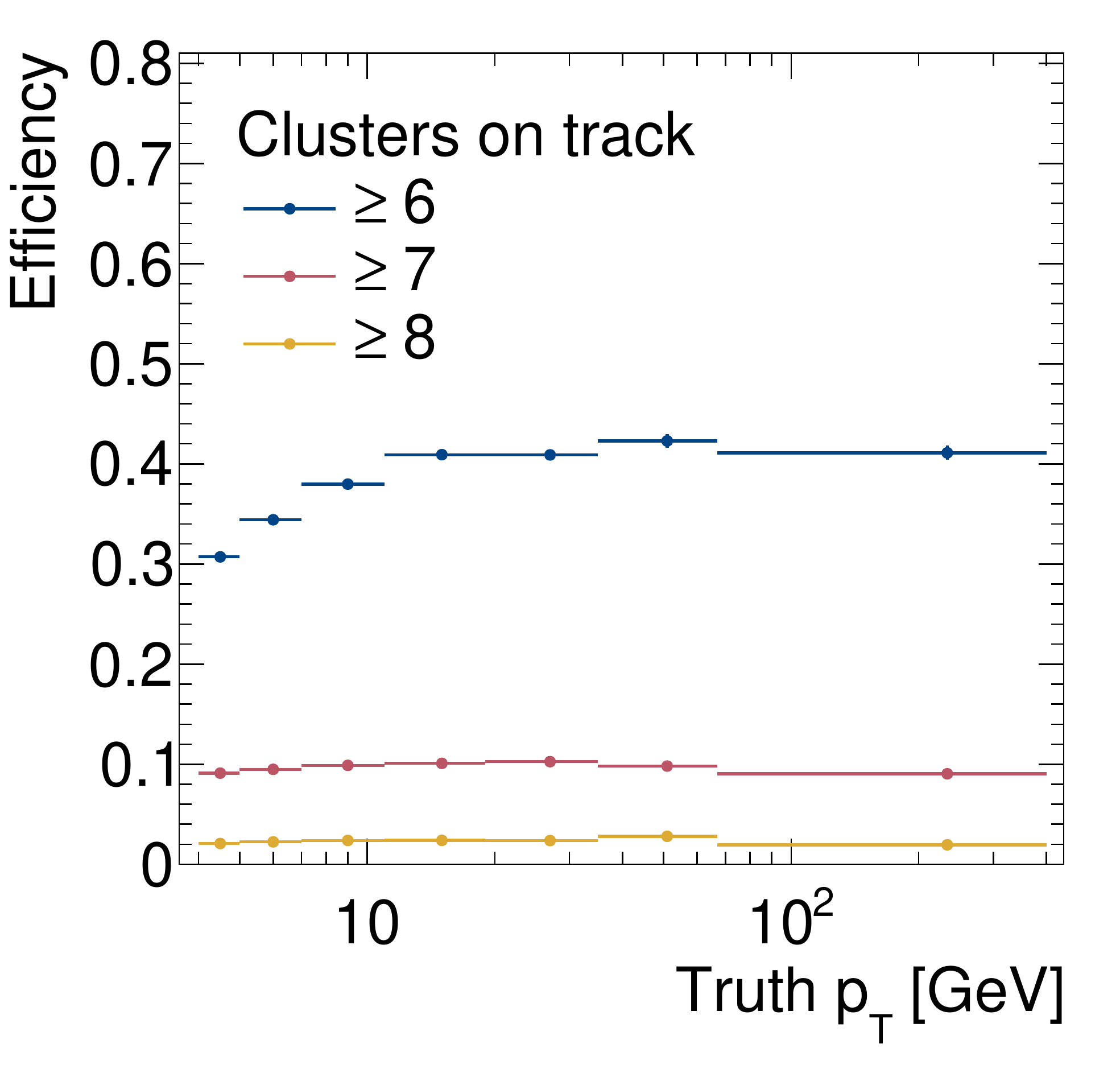}
  \end{minipage}
  \hspace{1em}
  \begin{minipage}{0.4\linewidth}
  \includegraphics[width=\columnwidth]{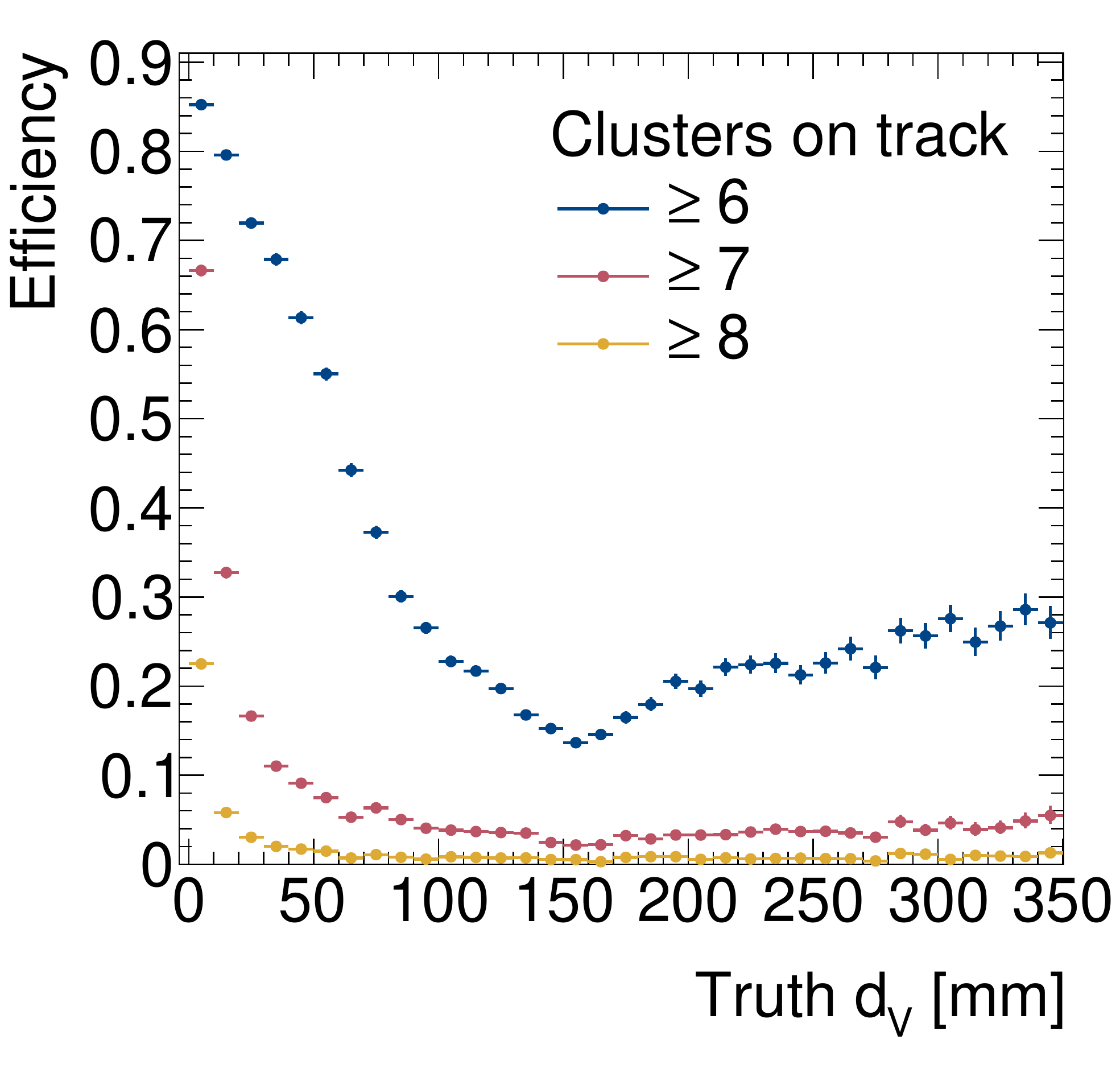}
  \end{minipage}
  \caption{Efficiency of finding at least 6, 7, or 8 hits in unique layers out of eight, as a function of $p_\text{T}$ (Left) and transverse vertex displacement (Right), for configuration 2, Hough transform method.}
  \label{fig:eff_63}
\end{figure}

Figure \ref{fig:eff_95} shows the efficiency as function of $p_\text{T}$ and
$d_\mathrm{V}$ of the truth track for the displaced-vertex sample with the
Hough transform configured for working point 1 (this is the mode optimized for
finding prompt muons from the beam spot). The efficiency is \SI{19}{\%} at
\SI{4}{\GeV}, peaks at \SIrange{10}{20}{\GeV} with \SI{27}{\%}, and decreases
to \SI{22}{\%} above \SI{67}{\GeV}. The efficiency as a function of
$d_\mathrm{V}$ starts at \SI{80}{\%} for $d_\mathrm{V}=\SI{0}{\mm}$ and
decreases to \SI{25}{\%} at $d_\mathrm{V}=\SI{50}{\mm}$; it also has a peculiar
shape with a minimum at \SI{150}{\mm}. Figure \ref{fig:eff_63} shows the same
plots when the Hough transform is configured for working point 2. The
efficiency increased with respect to working point 1, especially for
$d_\mathrm{V} < \SI{50}{\mm}$. It also plateaus at approximately \SI{40}{\%}
for $p_\text{T}>\SI{10}{\GeV}$, rather than decreasing at higher $p_\text{T}$.
The efficiencies for the other track parameters are found in Appendix~
\ref{app:hough_additional_distribution}.

\begin{figure}
  \centering
  \begin{minipage}{0.4\linewidth}
  \includegraphics[width=\columnwidth]{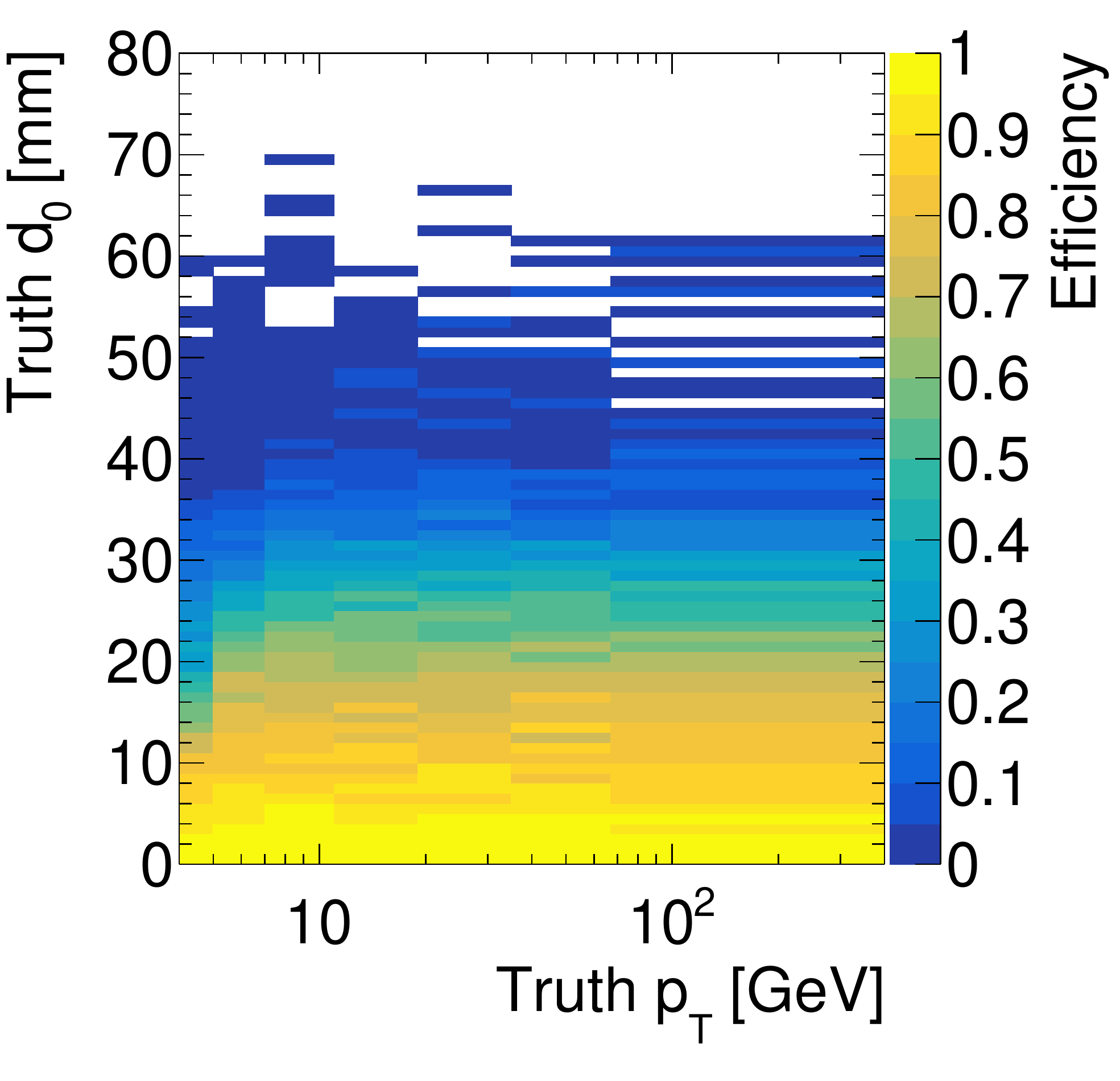}
  \end{minipage}
  \hspace{1em}
  \begin{minipage}{0.4\linewidth}
  \includegraphics[width=\columnwidth]{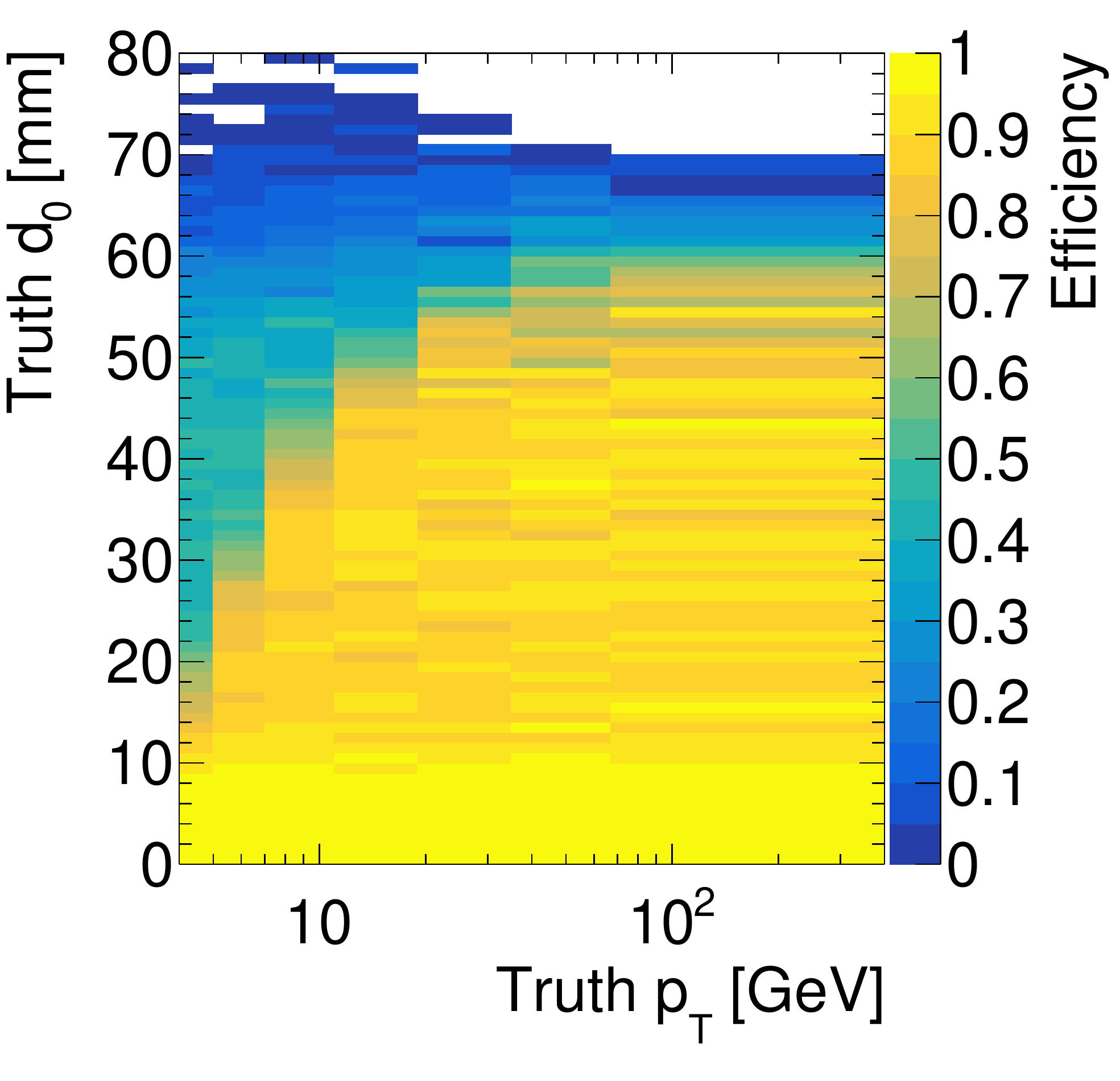}
  \end{minipage}
   \caption{Efficiency of finding at least 6  hits in unique layers out of 8 as a function of $p_\text{T}$ and $d_0$ for the Hough transform method. The left figure shows configuration 1 while the right figure shows configuration 2.}
  \label{fig:eff_2d}
\end{figure}

Figure~\ref{fig:eff_2d} shows two-dimensional efficiency distributions as a
function of the impact parameter ($d_0$) and $p_\mathrm{T}$ of the track. As
expected, the Hough transform has trouble finding tracks with low
$p_\mathrm{T}$ and high $d_0$.

\begin{figure}
  \centering
  \includegraphics[width=0.7\columnwidth]{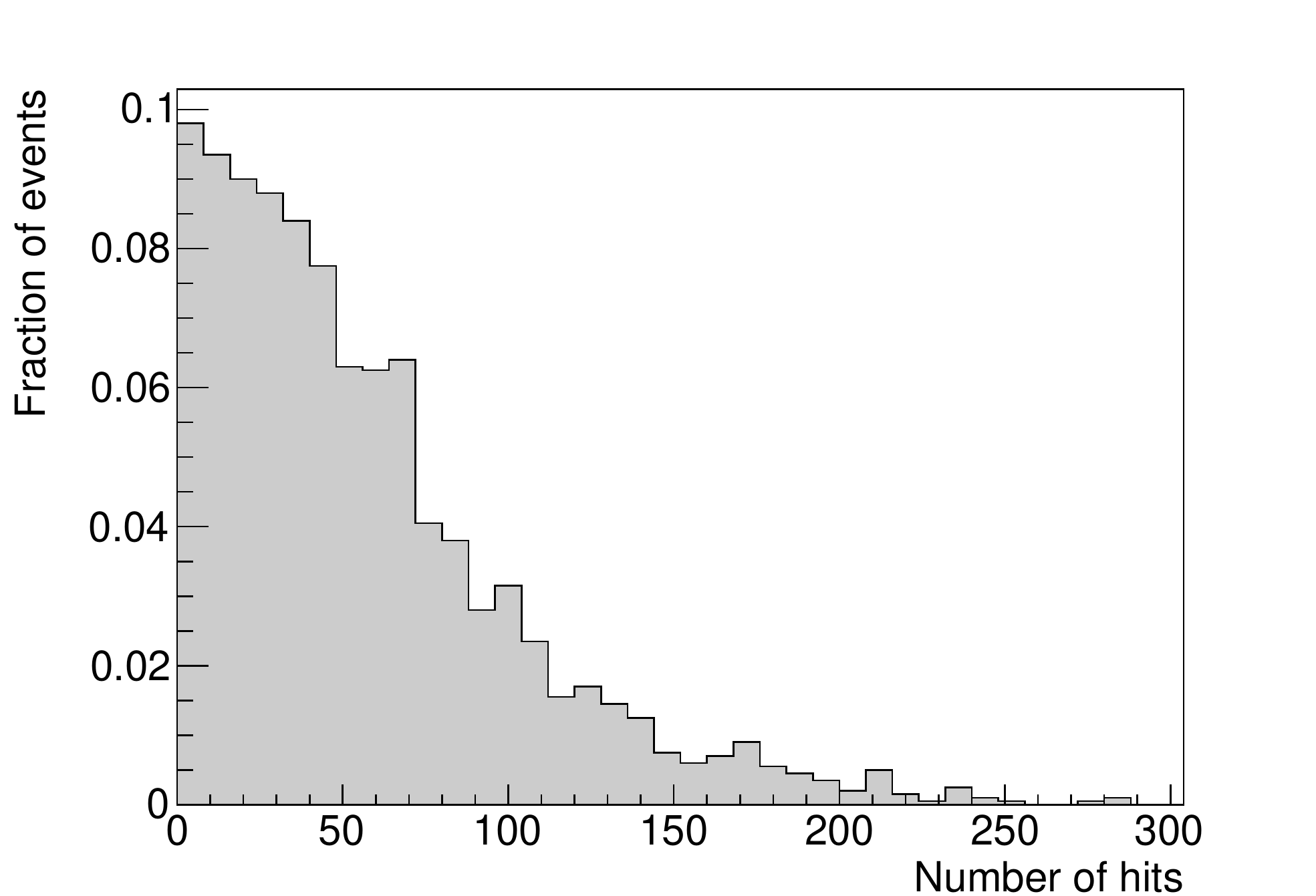}
  \caption{Number of hits passing the Hough transform for minimum bias events with a pile-up of 200.}
  \label{fig:nhitspassed_minbias}
\end{figure}

Track from soft QCD events generally have $p_\text{T}$ lower than the minimum of $\SI{4}{\GeV}$ that the Hough transform is configured for. This is the reason the Hough transform can suppress such backgrounds. Figure \ref{fig:nhitspassed_minbias} shows the distributions of the number of hits from minimum bias with pile-up 200 that pass the Hough transform selection for working point 1; the mean number of hist passing the Hough transform is 56.

\begin{figure}
  \centering
  \begin{minipage}{0.4\linewidth}
  \includegraphics[width=\columnwidth]{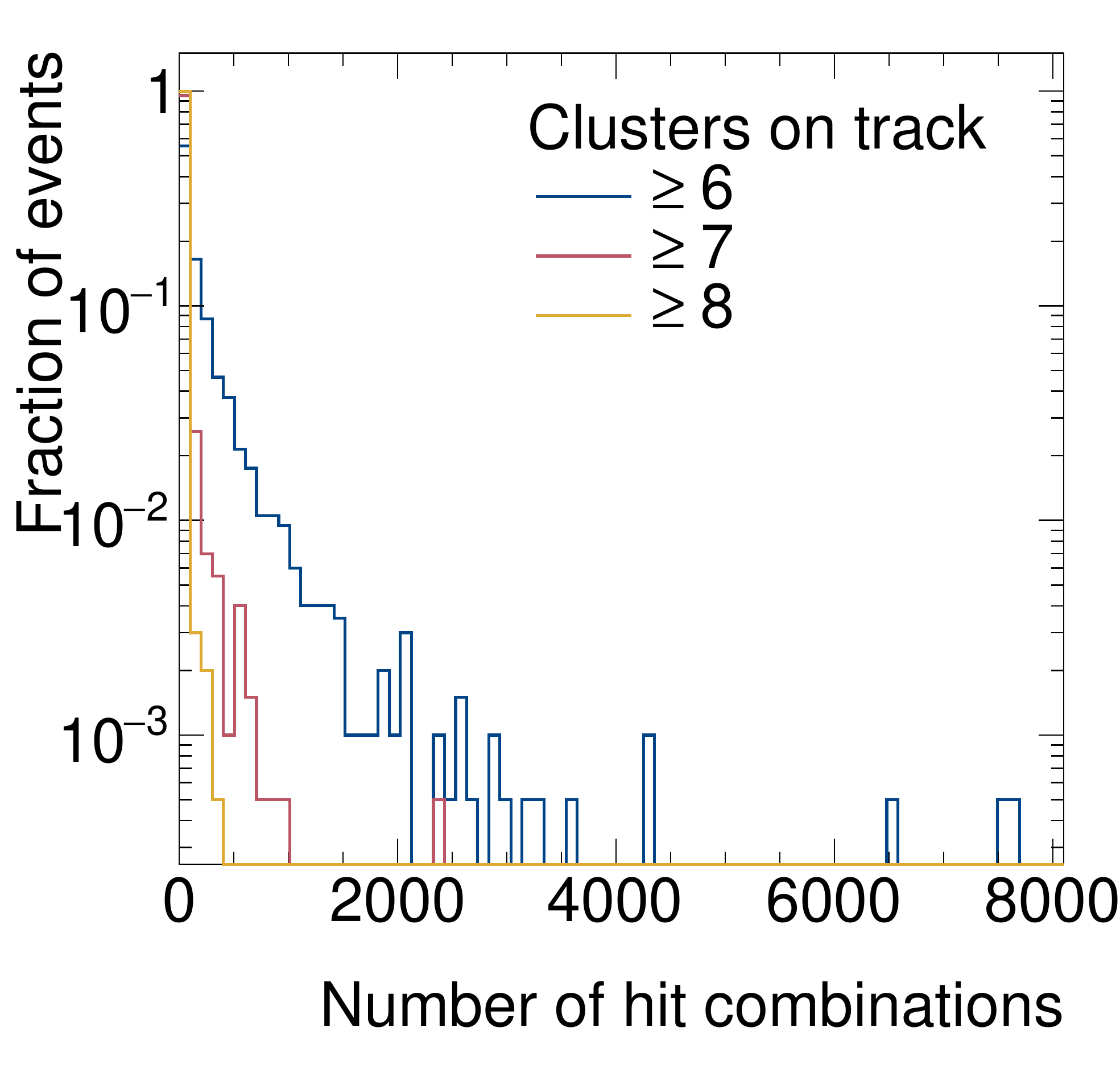}
  \end{minipage}
  \hspace{1em}
  \begin{minipage}{0.4\linewidth}
  \includegraphics[width=\columnwidth]{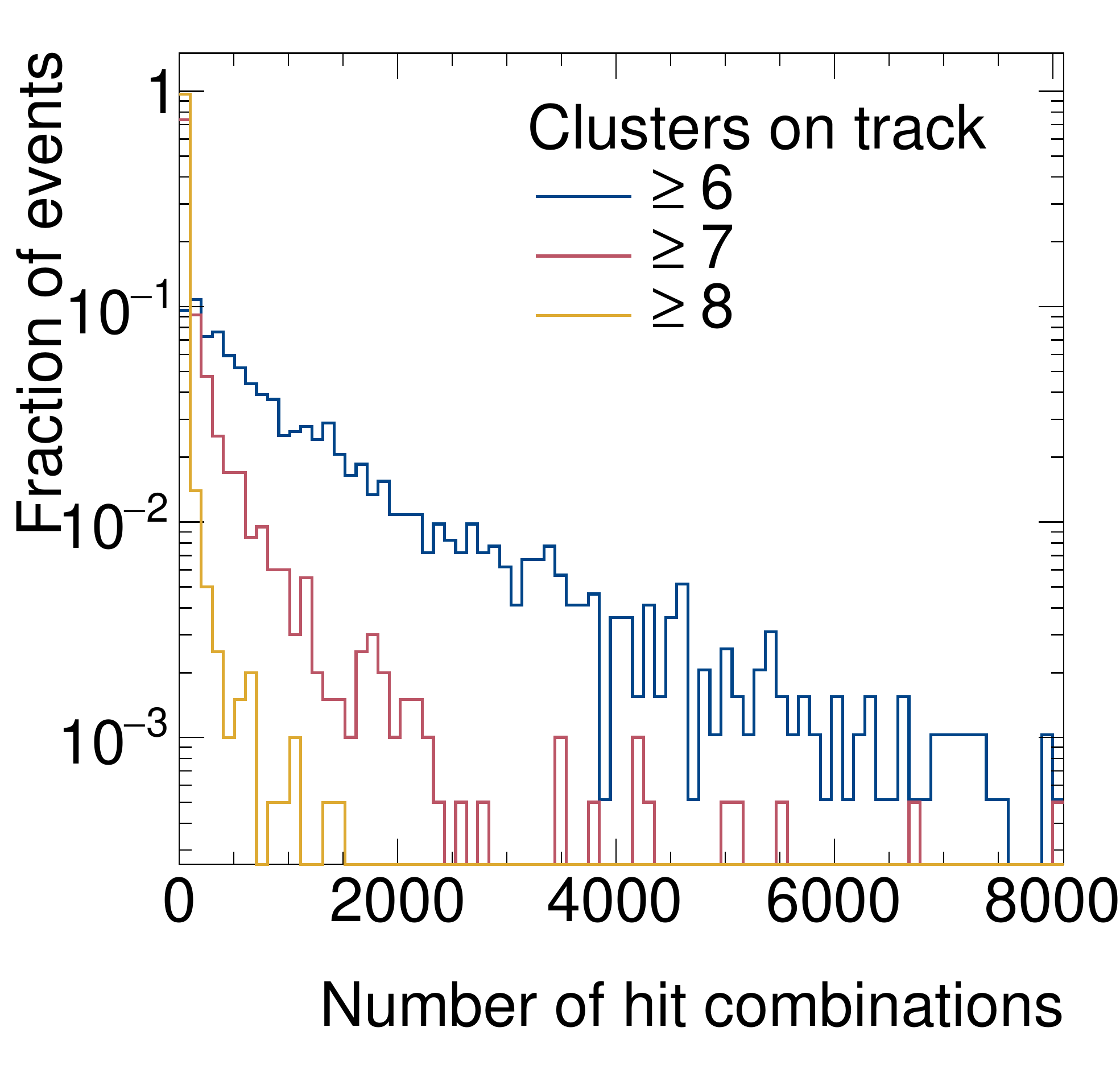}
  \end{minipage}
  \caption{Distribution of the number of possible hits combinations in a minimum bias sample with pile-up 200, requiring at least 6, 7, or 8 hits in unique layers out of 8, for configurations 1 (Left) and 2 (Right), Hough transform method.}
  \label{fig:hit_comb}
\end{figure}

The left panel of Figure \ref{fig:hit_comb} shows the number of hit
combinations in minimum bias, calculated according to equation
\ref{eq:cluster_combinations}, for the Hough transform using working point 1.
The mean is 84, but the distributions has a long tail with the
99\textsuperscript{th} percentile at 2017 combinations. The right panel of
Figure \ref{fig:hit_comb} shows the results for working point 2 that has 63
bins in $A\frac{qB}{p_\text{T}}$ compared to working point 1 that has 95
$A\frac{qB}{p_\text{T}}$ bins. The number of hit combinations increases to a
mean of 300 with 99\textsuperscript{th} percentile at 6530 combinations.

\begin{figure}
  \centering
  \begin{minipage}{0.4\linewidth}
  \includegraphics[width=\columnwidth]{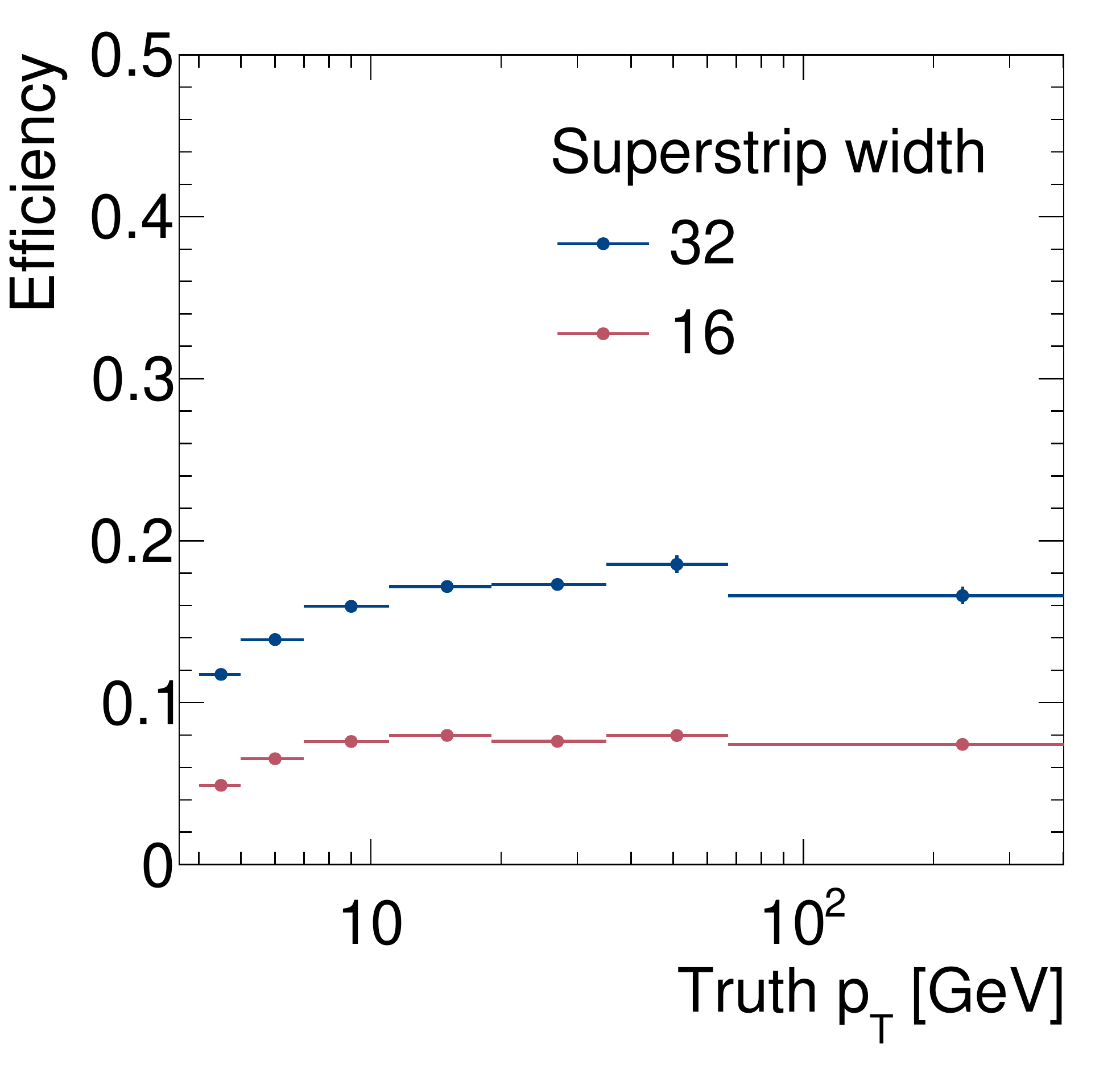}
  \end{minipage}
  \hspace{1em}
  \begin{minipage}{0.4\linewidth}
  \includegraphics[width=\columnwidth]{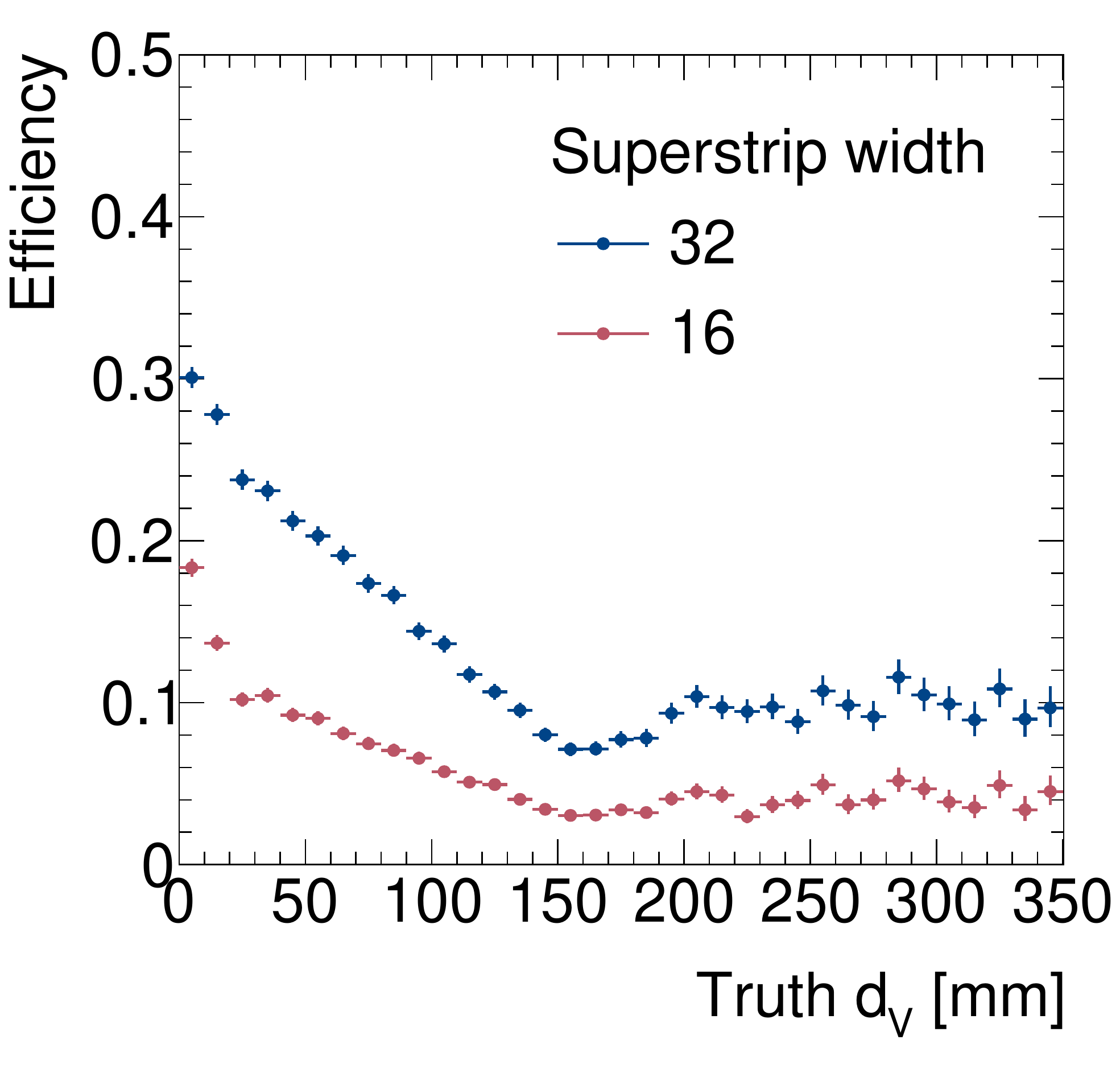}
  \end{minipage}
  \caption{Efficiency for tracks from displaced vertex with pattern banks trained on prompt muons from beam spot for SSW of 16 and 32 as a function of $p_\text{T}$ (Left) and transverse vertex displacement (Right).}
  \label{fig:eff_ssw16_32}
\end{figure}

Figure \ref{fig:eff_ssw16_32} shows the efficiency as function of $p_\text{T}$
and $d_V$ of the truth track for the displaced-vertex sample with the pattern
matching method when using a pattern bank trained on muons originating from the
beam spot, that is: not trained for tracks from displaced vertex, which is the
default system. The pattern banks are trained to give an efficiency above
$\SI{99}{\%}$ with a minimum of 6 matched hits in an 8 hit pattern. The study shows
about \SI{4}{\%} efficiency for SSW of 16 for finding tracks from muons with
displaced vertex above $\SI{150}{\mm}$, with efficiency increasing for low
$d_V$. The corresponding number for SSW of 32 is around \SI{10}{\%}.
The efficiency is expected to approach \SI{99}{\%} for $d_V$ approaching \SI{0}{\milli\meter},
however, Figure~\ref{fig:eff_ssw16_32} shows an efficiency of \SI{\sim30}{\%} for tracks using
a SSW of 32. The prompt muon tracks used to train the pattern banks used in Figure~\ref{fig:eff_ssw16_32}
are generated from flat distributions in $0.1 < \eta < 0.3$ and $0.3 < \phi < 0.5$. This is
in contrast with how the displaced tracks are generated, where the direction depends
on the vertex position as discussed in Section~\ref{sec:simulation}. Restricting the
displaced tracks to the same region as the prompt tracks recovers the efficiency as
further demonstrated in Appendix~\ref{app:pattern_additional_distribution}.

\begin{figure}
  \centering
  \begin{minipage}{0.4\linewidth}
  \includegraphics[width=\columnwidth]{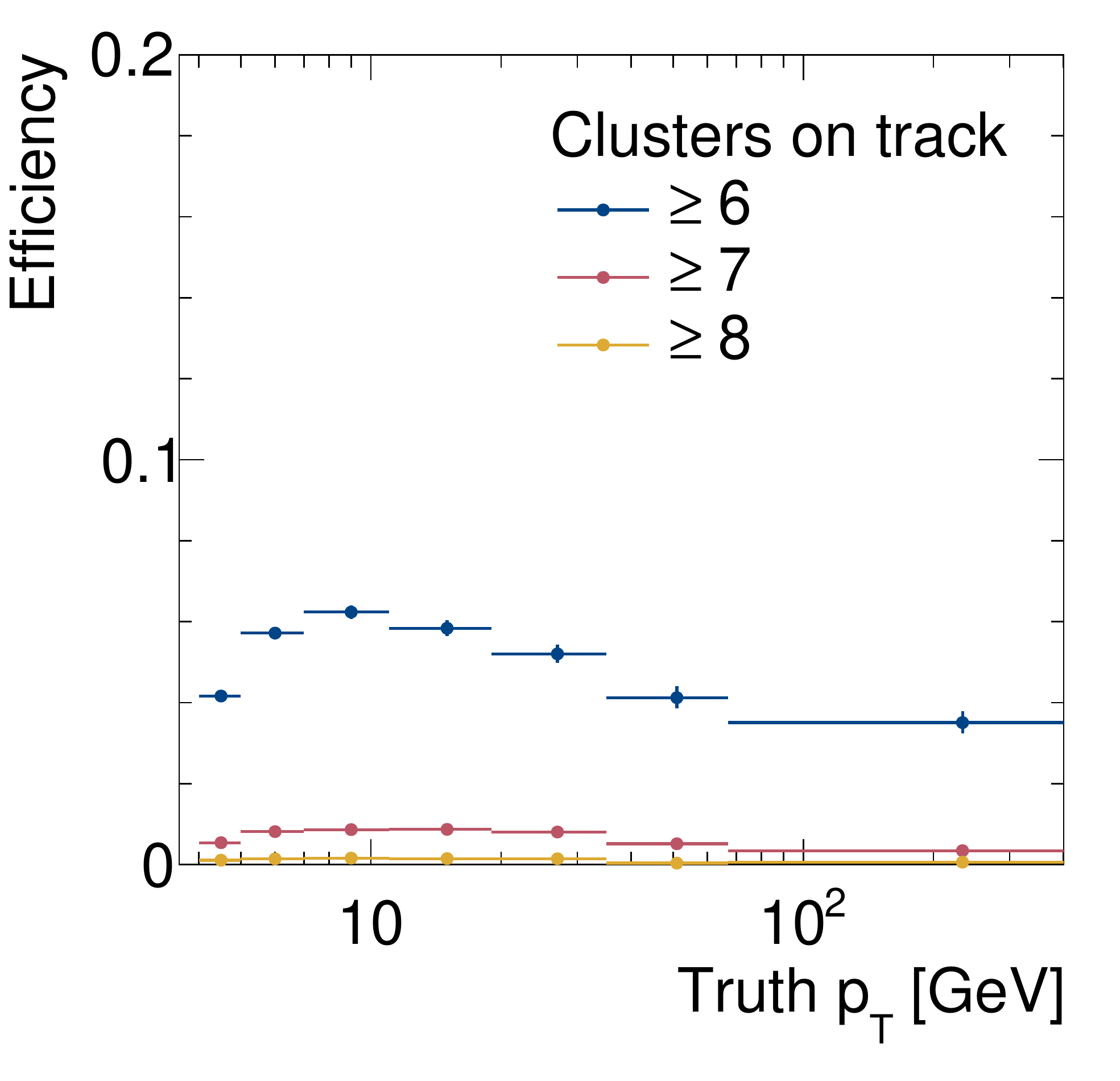}
  \end{minipage}
  \hspace{1em}
  \begin{minipage}{0.4\linewidth}
  \includegraphics[width=\columnwidth]{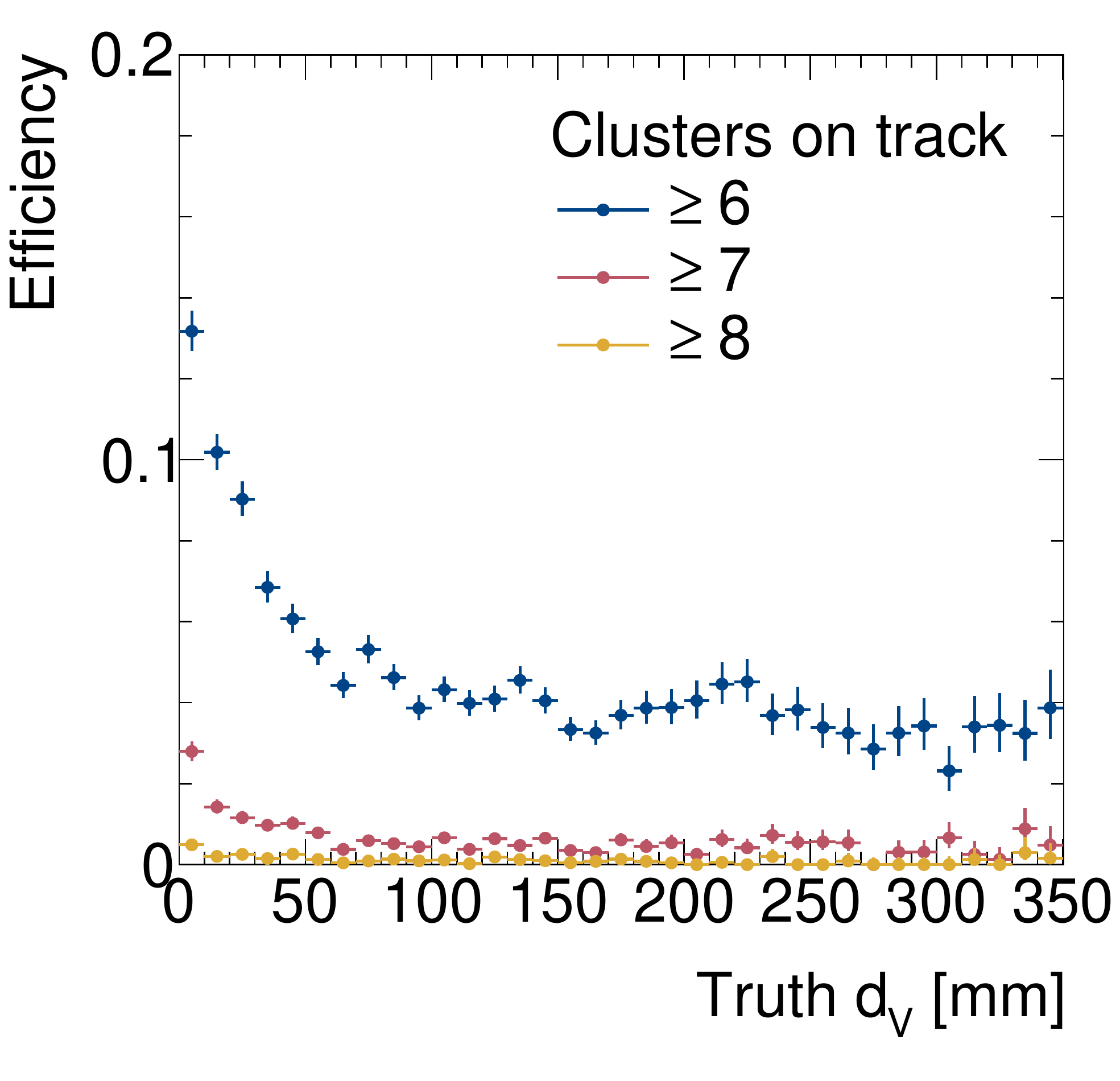}
  \end{minipage}
  \caption{Efficiency for displaced tracks as a function of $p_\text{T}$ (Left) and transverse vertex displacement (Right), using a dedicated pattern bank trained for displaced tracks with 100k patterns, and a SSW of 16, pattern matching method.}
  \label{fig:eff_100k_ssw16}
\end{figure}

\begin{figure}
  \centering
  \begin{minipage}{0.4\linewidth}
  \includegraphics[width=\columnwidth]{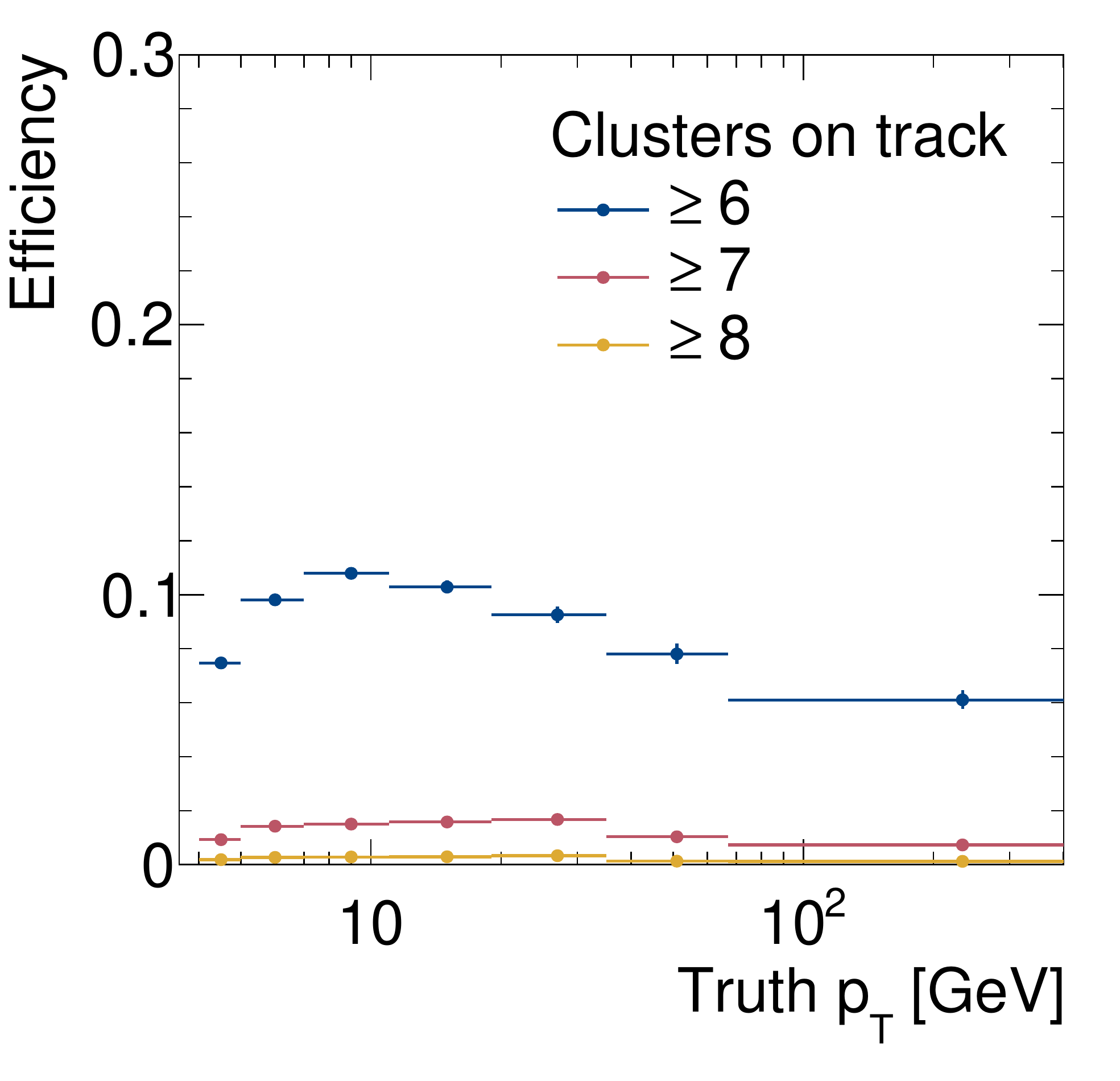}
  \end{minipage}
  \hspace{1em}
  \begin{minipage}{0.4\linewidth}
  \includegraphics[width=\columnwidth]{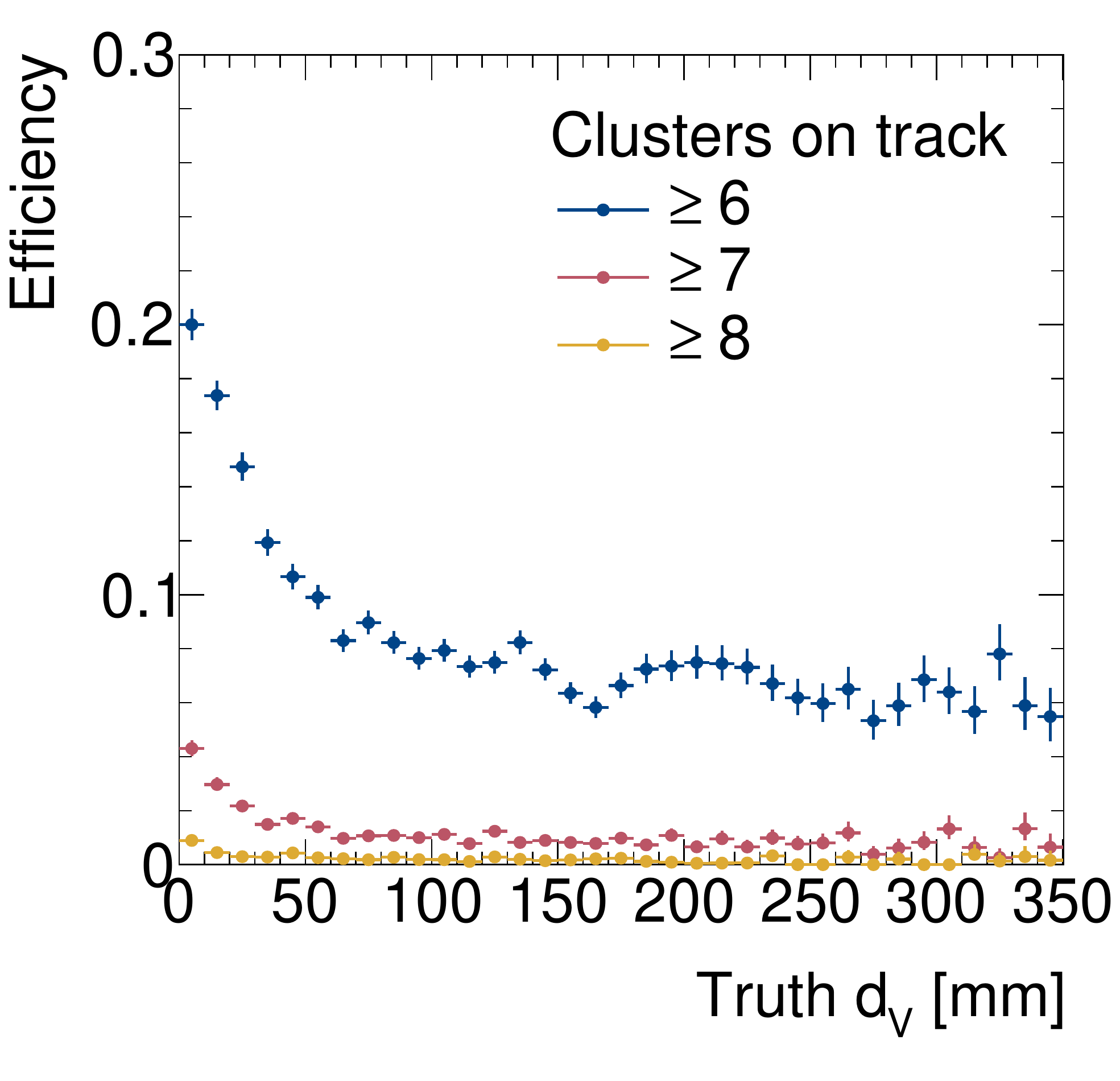}
  \end{minipage}
  \caption{Efficiency for displaced tracks as a function of $p_\text{T}$ (Left) and transverse vertex displacement (Right), using a dedicated pattern bank trained for displaced tracks with 200k patterns, and a SSW of 16, pattern matching method.}
  \label{fig:eff_200k_ssw16}
\end{figure}

\begin{figure}
  \centering
  \begin{minipage}{0.4\linewidth}
  \includegraphics[width=\columnwidth]{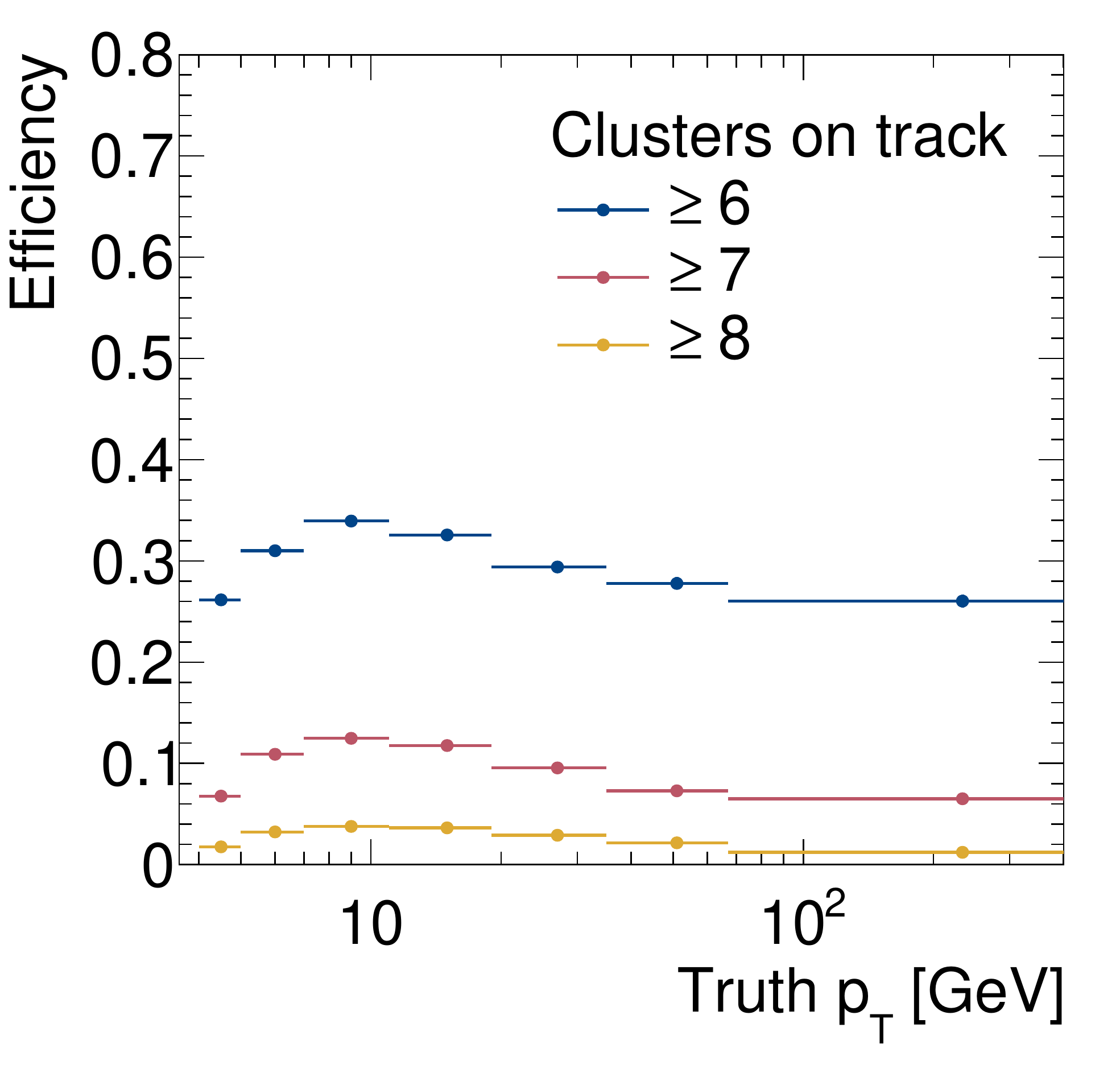}
  \end{minipage}
  \hspace{1em}
  \begin{minipage}{0.4\linewidth}
  \includegraphics[width=\columnwidth]{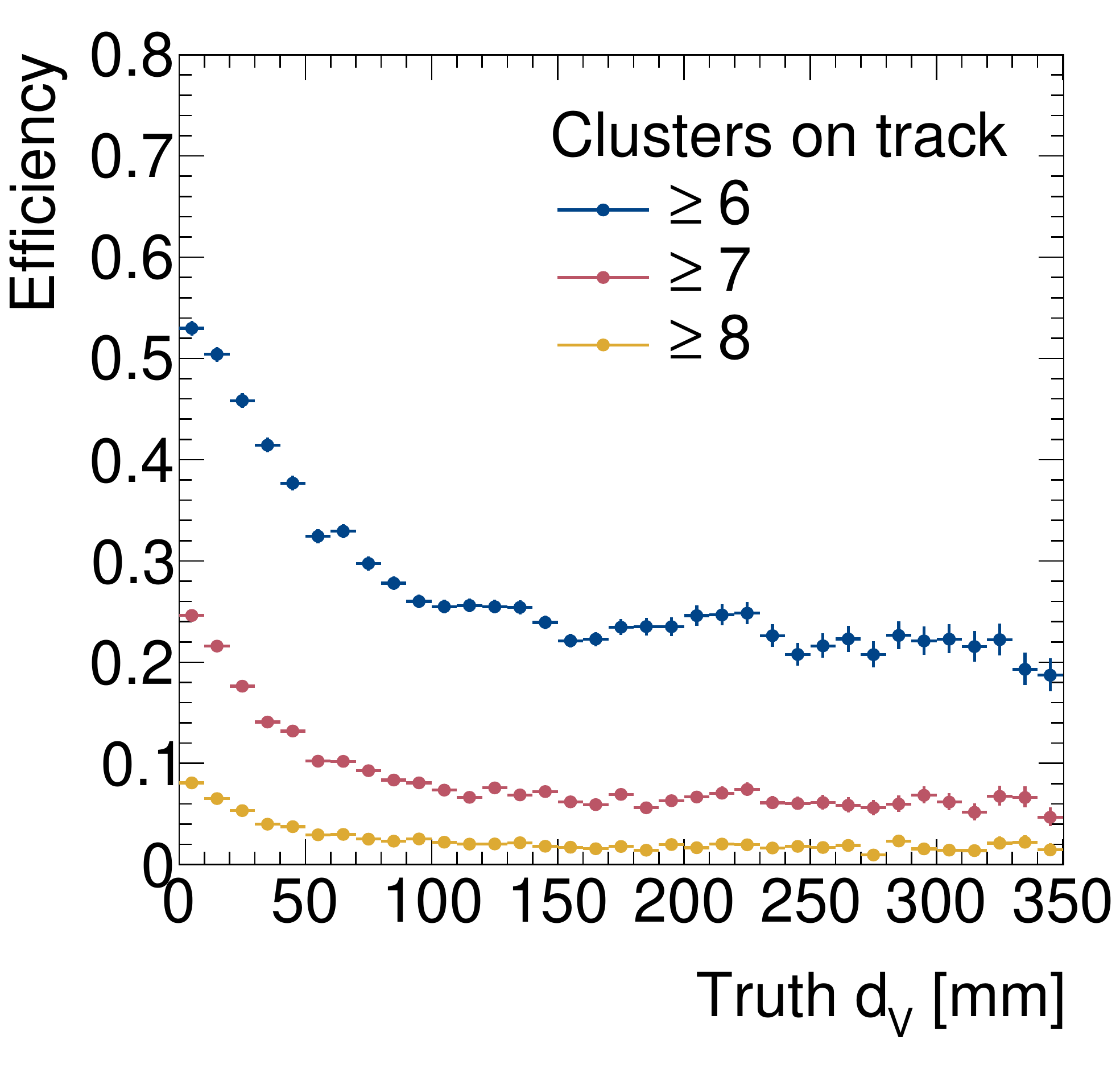}
  \end{minipage}
  \caption{Efficiency for displaced tracks as a function of $p_\text{T}$ (Left) and transverse vertex displacement (Right), using a dedicated pattern bank trained for displaced tracks with 200k patterns, and a SSW of 32, pattern matching method.}
  \label{fig:eff_200k_ssw32}
\end{figure}

Figure \ref{fig:eff_100k_ssw16} shows the efficiency as function of
$p_\text{T}$ and $d_V$ of the truth track for the displaced-vertex sample with
the pattern matching method when using a dedicated pattern bank trained for
displaced tracks with \SI{100}{\kilo\nothing} patterns, and a SSW of 16. The
efficiency is around \SI{10}{\%} at $d_V=\SI{20}{\mm}$ and stays flat at about
\SI{4}{\%} from $d_V = \SI{50}{\mm}$. By doubling the size of the pattern bank
this efficiency becomes higher: above \SI{8}{\%} for the whole $p_\text{T}$
spectrum, peaking at \SI{10}{\GeV}, and reaching \SI{20}{\%} for low values of
$d_V$, as shown in Figure~\ref{fig:eff_200k_ssw16}.
Figure~\ref{fig:eff_200k_ssw32} shows the efficiency when using a pattern bank
of \SI{200}{\kilo\nothing} patterns and a SSW width of 32. In this
configuration, the efficiency stays well above \SI{20}{\%} overall, reaching
more than \SI{50}{\%} of efficiency for low values of $d_V = \SI{50}{\mm}$.

\begin{figure}
  \centering
  \begin{minipage}{0.4\linewidth}
  \includegraphics[width=\columnwidth]{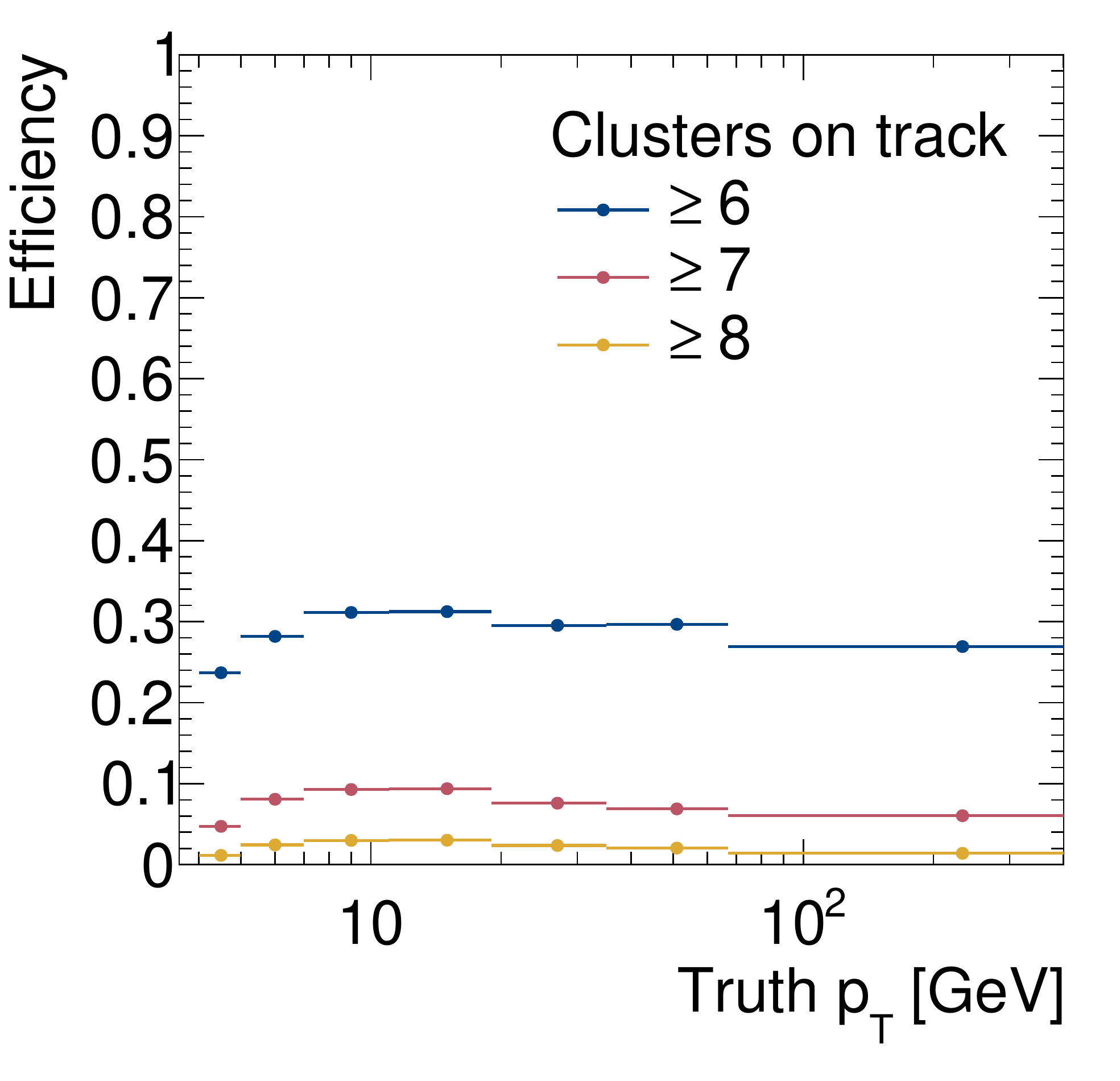}
  \end{minipage}
  \hspace{1em}
  \begin{minipage}{0.4\linewidth}
  \includegraphics[width=\columnwidth]{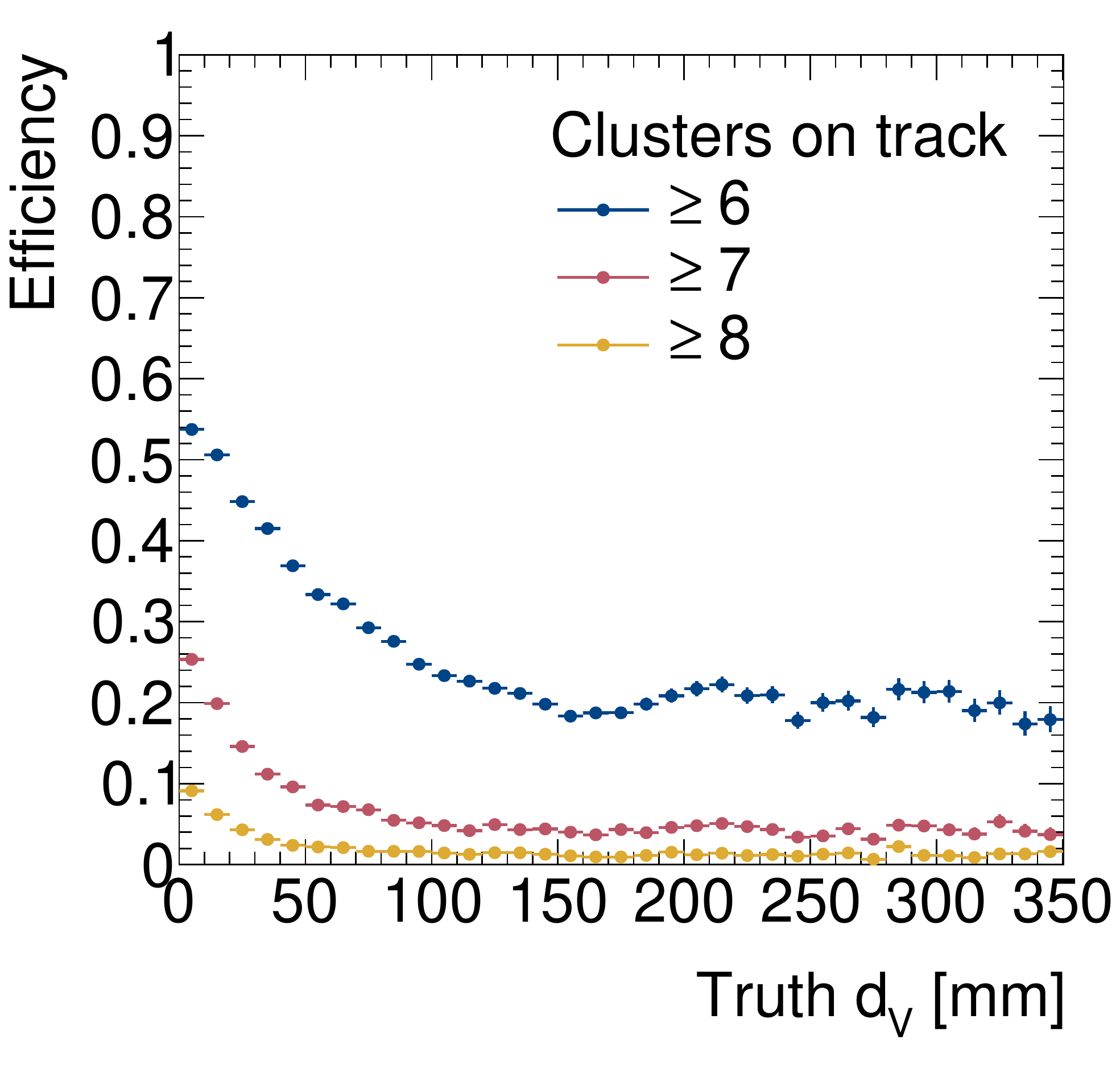}
  \end{minipage}
  \caption{Efficiency for displaced tracks as a function of $p_\text{T}$ (Left) and transverse vertex displacement (Right), using a mixed pattern bank with 1M patterns from which 100k correspond to displaced patters, using a SSW of 32, pattern matching method.}
  \label{fig:mix10}
\end{figure}

\begin{figure}
  \centering
  \begin{minipage}{0.4\linewidth}
  \includegraphics[width=\columnwidth]{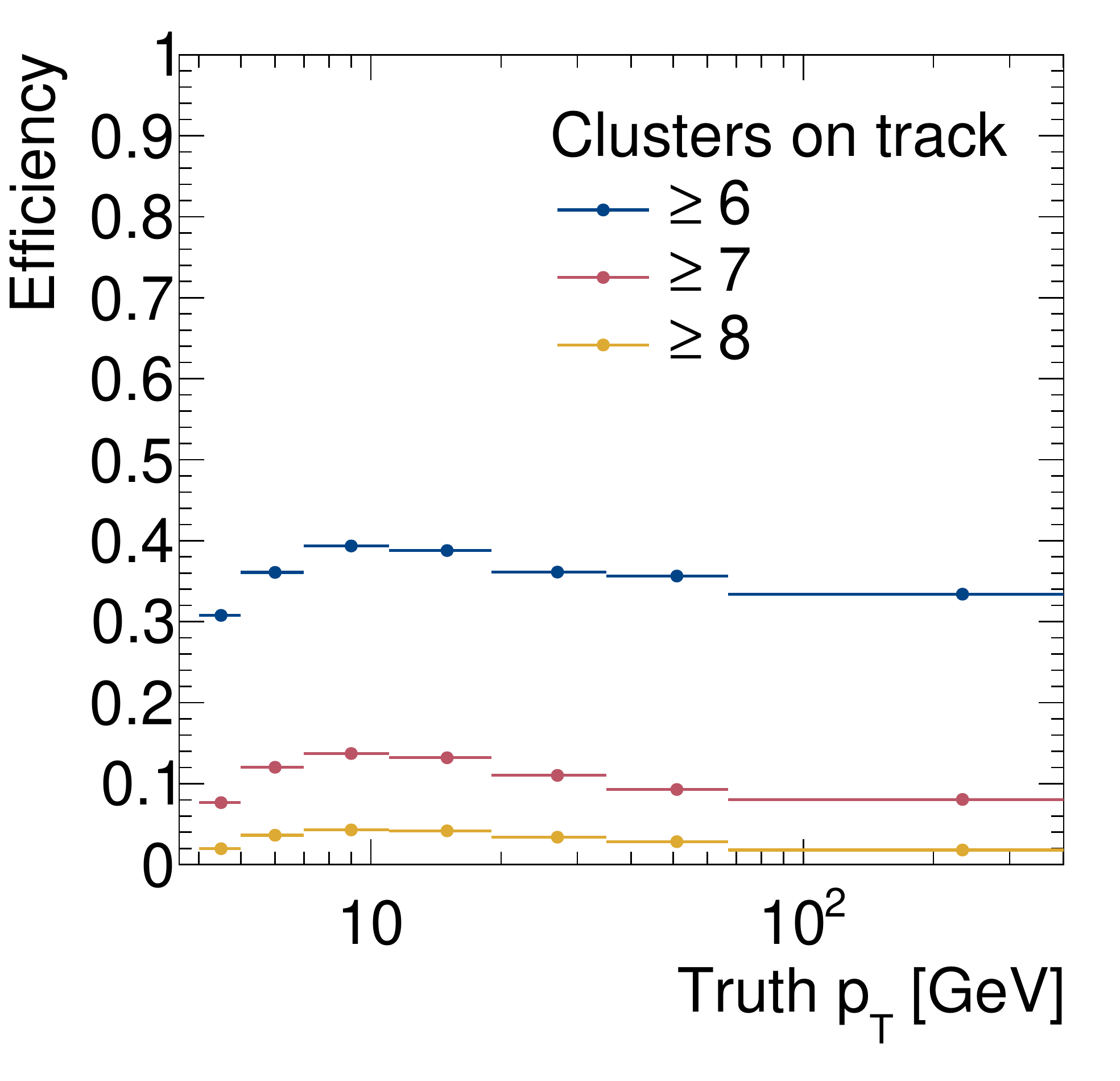}
  \end{minipage}
  \hspace{1em}
  \begin{minipage}{0.4\linewidth}
  \includegraphics[width=\columnwidth]{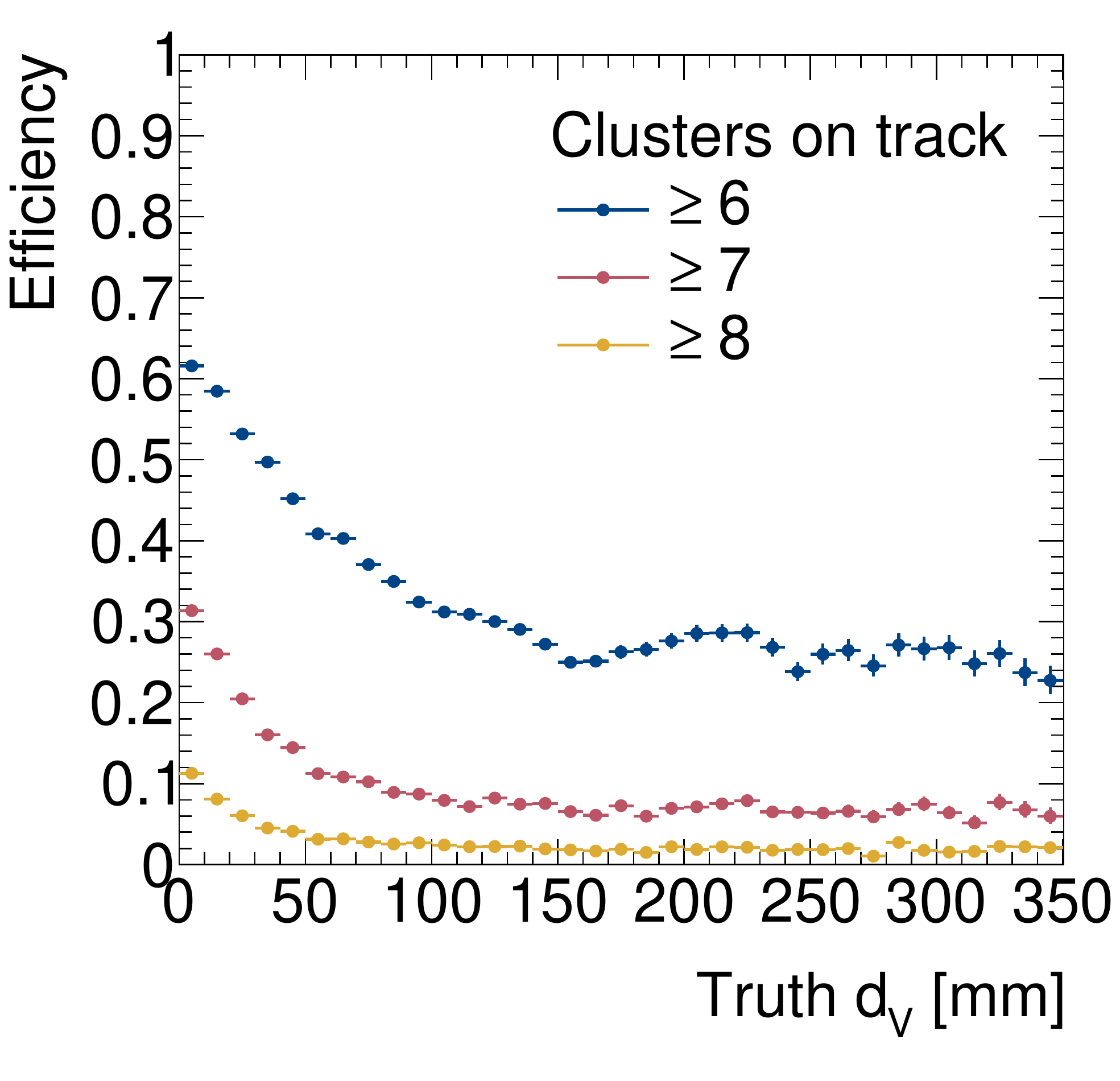}
  \end{minipage}
  \caption{Efficiency for displaced tracks as a function of $p_\text{T}$ (Left) and transverse vertex displacement (Right), using a mixed pattern bank with 1M patterns from which 200k correspond to displaced patters, using a SSW of 32, pattern matching method.}
  \label{fig:mix20}
\end{figure}

\begin{figure}
  \centering
  \begin{minipage}{0.4\linewidth}
  \includegraphics[width=\columnwidth]{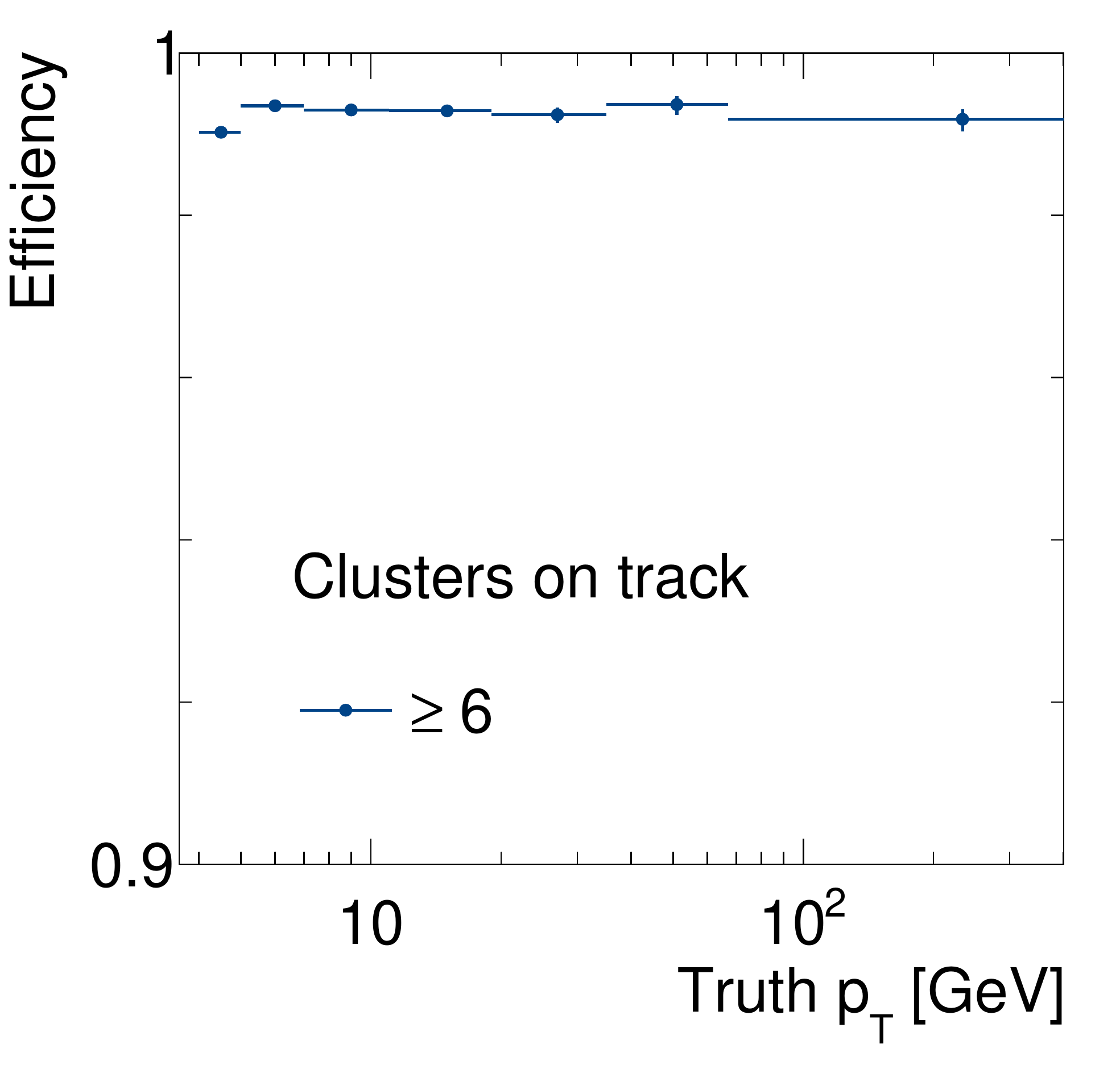}
  \end{minipage}
  \hspace{1em}
  \begin{minipage}{0.4\linewidth}
  \includegraphics[width=\columnwidth]{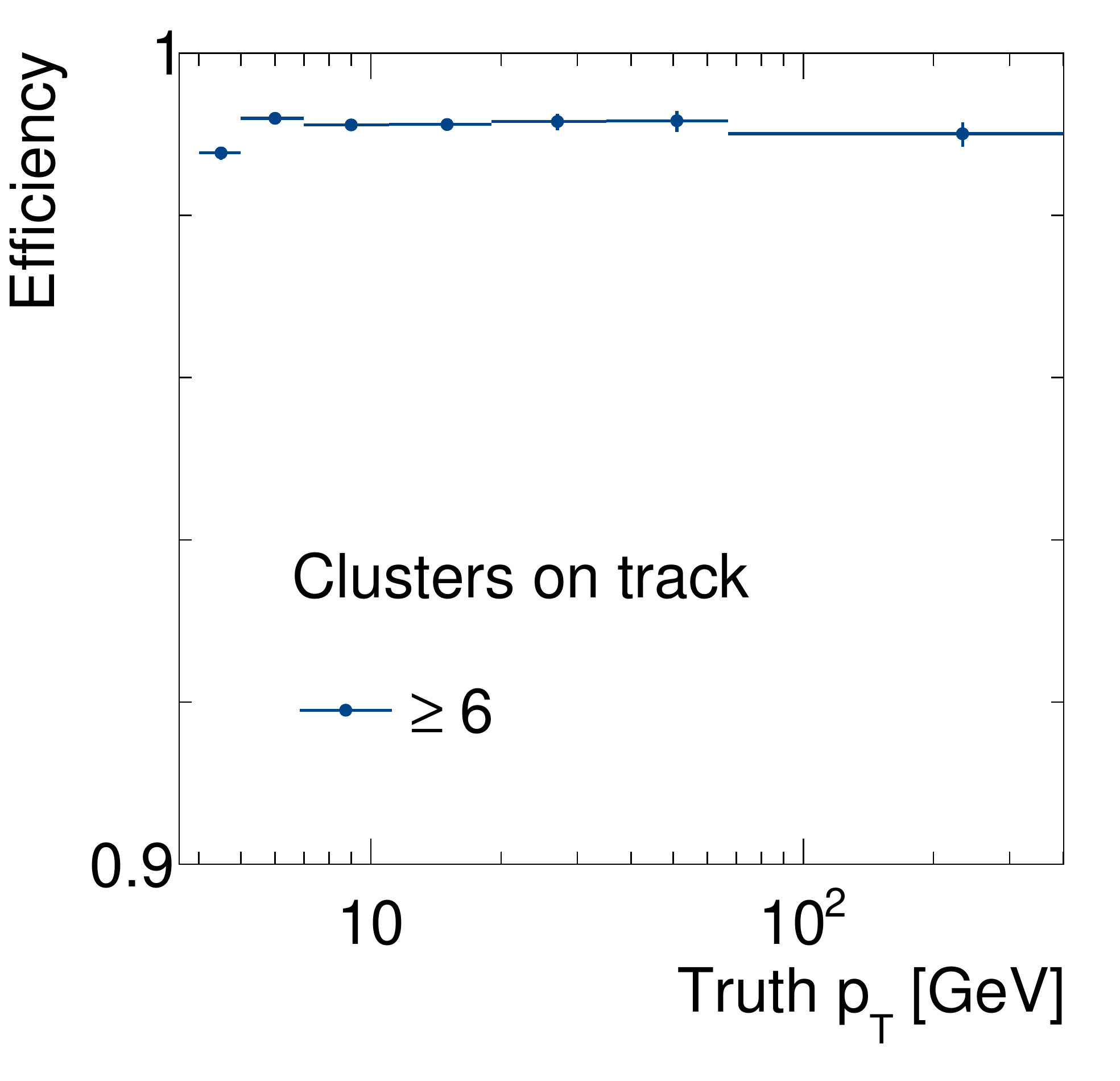}
  \end{minipage}
  \caption{Efficiency for prompt tracks using a mixed pattern bank with 1M patterns from which 100k (Left) or 200k (Right) correspond to displaced patterns, using a SSW of 32, pattern matching method.}
  \label{fig:mix20_prompt}
\end{figure}

In a realistic scenario, the budget of patterns to be stored will have to be
shared between prompt and displaced patterns. Figure~\ref{fig:mix10} shows the
efficiency for displaced tracks using a mixed pattern bank with
\SI{1}{\mega\nothing} patterns where \SI{100}{\kilo\nothing} correspond to
displaced signatures and \SI{900}{\kilo\nothing} correspond to prompt tracks.
The efficiency stays well above \SI{20}{\%} for all values of $d_V$ and peaks
at  \SI{30}{\%} for relatively low $p_\text{T}$  tracks of
\SIrange{10}{20}{\GeV}. A mixed pattern bank with a higher number of patterns
trained on displaced tracks, for example \SI{20}{\%}, as shown in
Figure~\ref{fig:mix20} would perform even better, with almost flat efficiencies
around \SI{30}{\%} for high values of $d_V$ and small values of $p_\text{T}$.
To dedicate more than \SI{10}{\%} of the pattern banks to displaced tracks is
something that could be considered in practice since it does not seem to
affect, in a significant way, the efficiency for prompt tracks. This is evident from the results presented in
Figure~\ref{fig:mix20_prompt}, where the efficiency as a function of
$p_\text{T}$ for prompt tracks stays above \SI{99}{\%} for the full range.

\begin{figure}
  \centering
  \includegraphics[width=0.4\columnwidth]{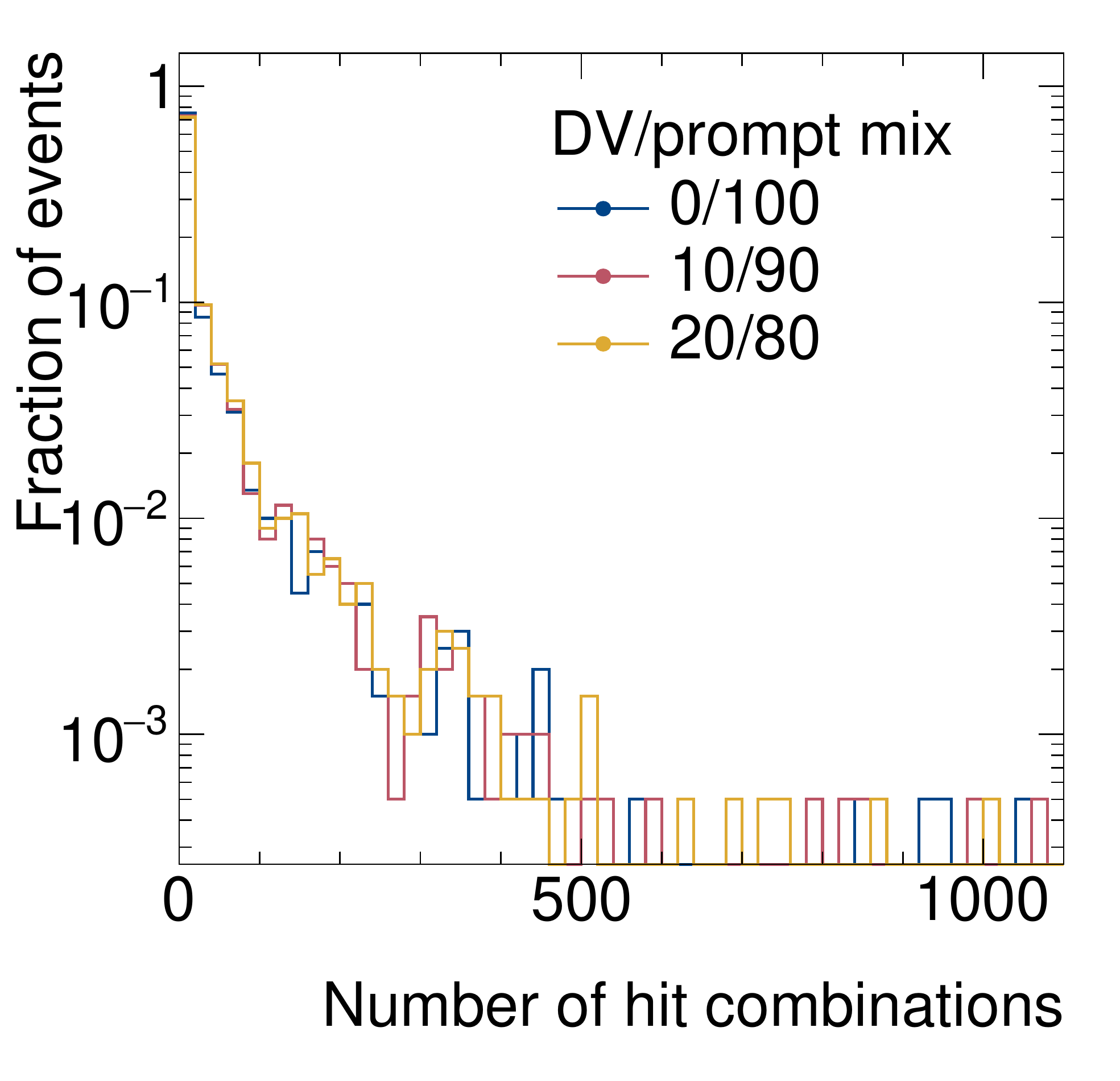}
  \caption{Distribution of the number of possible hits combinations in a minimum bias sample with pile-up 200, using a pattern bank of 1M patterns trained, using a SSW of 32, with: 10\% displaced tracks (red), 20\% displaced tracks (yellow), and only prompt tracks (blue). Pattern matching method.}
  \label{fig:hit_comb_pm_mix}
\end{figure}

\begin{figure}
  \centering
  \begin{minipage}{0.4\linewidth}
      \includegraphics[width=\columnwidth]{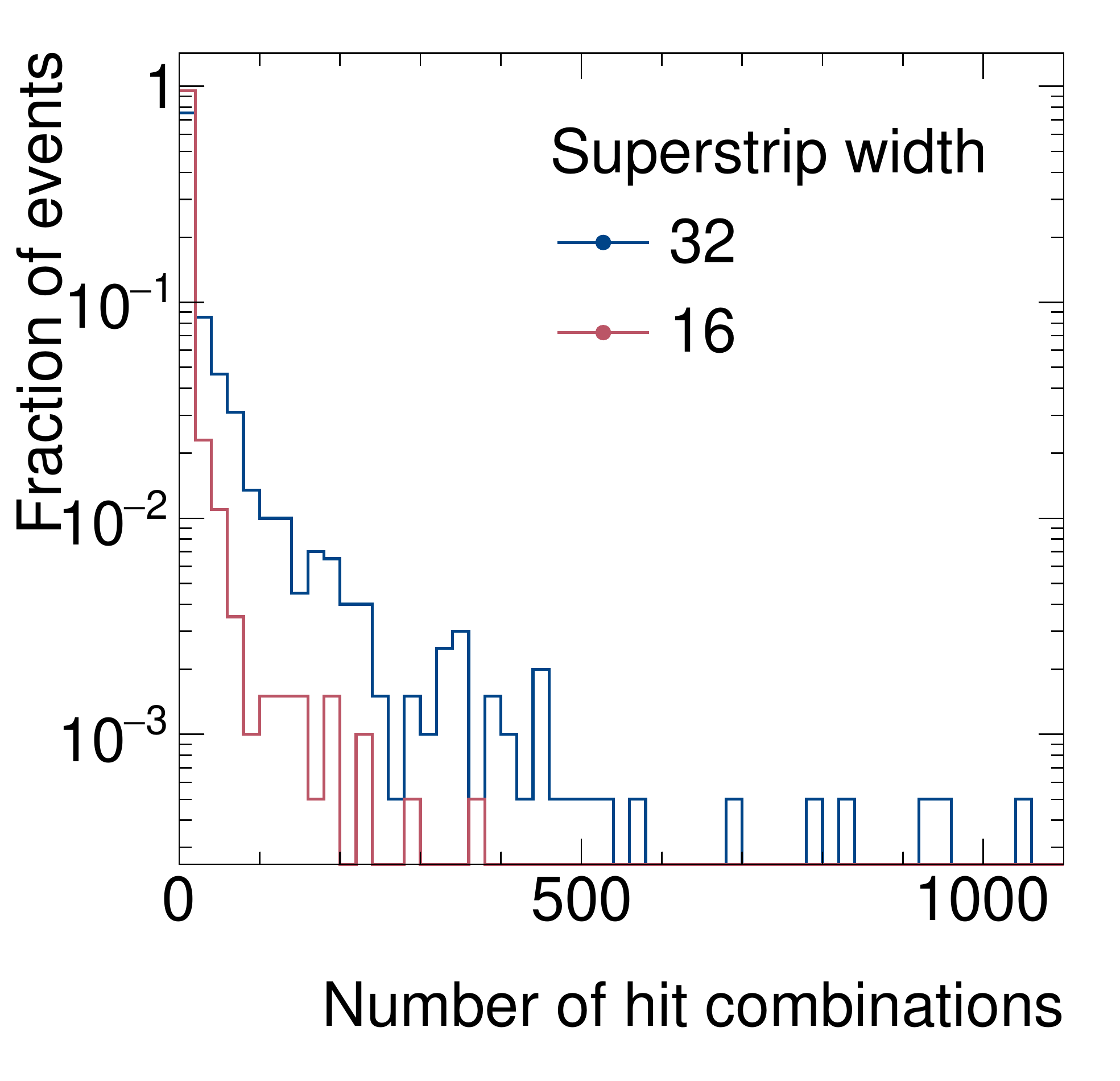}
  \end{minipage}
  \hspace{1em}
  \begin{minipage}{0.4\linewidth}
  \includegraphics[width=\columnwidth]{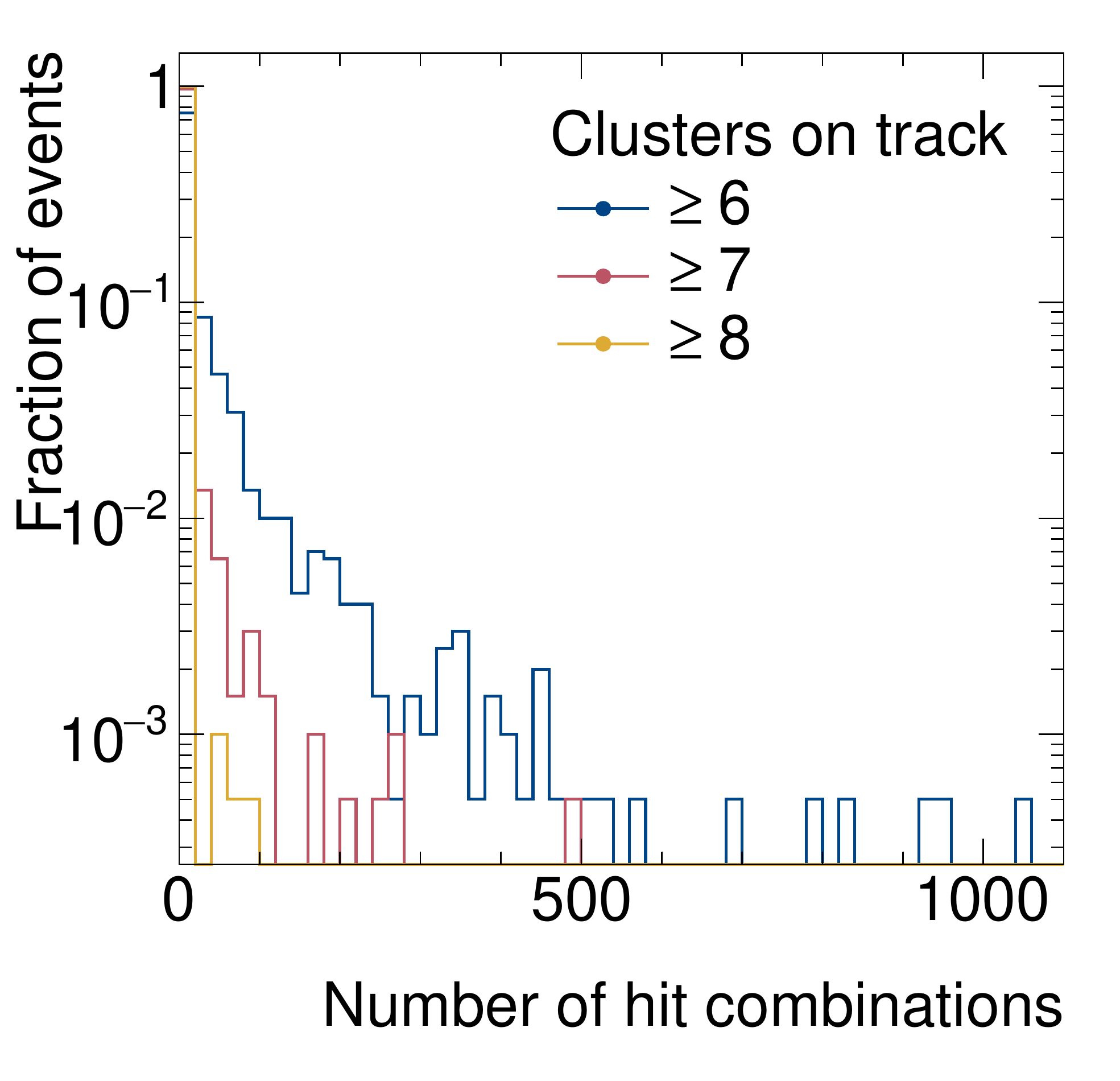}
  \end{minipage}
  \caption{ Distribution of the number of possible hits combinations in a minimum bias sample with pile-up 200, for 16 SSW and 32 SSW (Left);  and requiring at least 6, 7, or 8 hits in unique layers out of 8, with a SSW of 32 (Right). Pattern matching method using a pattern bank trained only with prompt tracks.}
  \label{fig:hitspr}
\end{figure}

Finally, we can also check the number of hit combinations in minimum bias for
the pattern matching method, calculated in the same way as for the Hough
transform. Figure \ref{fig:hit_comb_pm_mix} shows the distribution of the
number of possible hits combinations in a minimum bias sample with pile-up 200,
using a pattern bank of 1M patterns trained with 10\%  and 20\% displaced
tracks and only prompt tracks. In Figure~\ref{fig:hitspr} we can see the number
of hit combinations when using 16 SSW compared with 32 SSW (Left), as well as
the number of hit combinations when requiring at least 6, 7, or 8 hits in
unique layers out of 8 (Right), on a pattern bank trained with only prompt
tracks.

\section{Summary and discussion}

We have studied with full detector simulation two methods that can be implemented in hardware-based processing, such as the proposed hardware track trigger for the Phase-II upgrade of the ATLAS detector for the HL-LHC~\cite{ATLAS-Phase2}, to identify displaced tracks. The first method, based on the Hough transform, shows promising efficiencies: above \SI{20}{\%} for displaced tracks with very high transverse vertex displacement, maintaining a flat efficiency around \SI{40}{\%} for $p_\text{T}>\SI{10}{\GeV}$. The second method, based on pattern recognition, tests the possibility of modifying the regular pattern banks trained for prompt muons produced in the central interaction region by including \SI{10}{\%} and \SI{20}{\%} extra patterns trained with muons originating from displaced vertices inside the tracker. The efficiency of the pattern matching is above \SI{30}{\%} for displaced tracks while maintaining above \SI{99}{\%} efficiency for prompt tracks.

Both methods will increase the load on the data acquisition system because of additional hits from minimum bias. However, this effect can be kept small if only events of interest are sent to the hardware-based processing for long-lived particles, how this can be done for the two methods here discussed is left for a future study.

This is a preliminary study that anticipates a promising performance of dedicated trigger strategies for the search for long-lived particles at the HL-LHC.  With small additional resources, the sensitivity for long-lived particles can be improved. We propose that such resources are reserved in the design of the future tracker triggers for the HL-LHC.

\acknowledgments
The Swedish Research Council supports Rebeca Gonzalez Suarez (VR 2017-05092) and Richard Brenner (VR 2015-04955).
\clearpage
\bibliographystyle{JHEP}
\bibliography{references}
\clearpage
\appendix

\section{Additional efficiency distributions for the Hough transform}
\label{app:hough_additional_distribution}

\begin{figure}[!ht]
  \centering
  \begin{minipage}{0.4\linewidth}
  \includegraphics[width=\columnwidth]{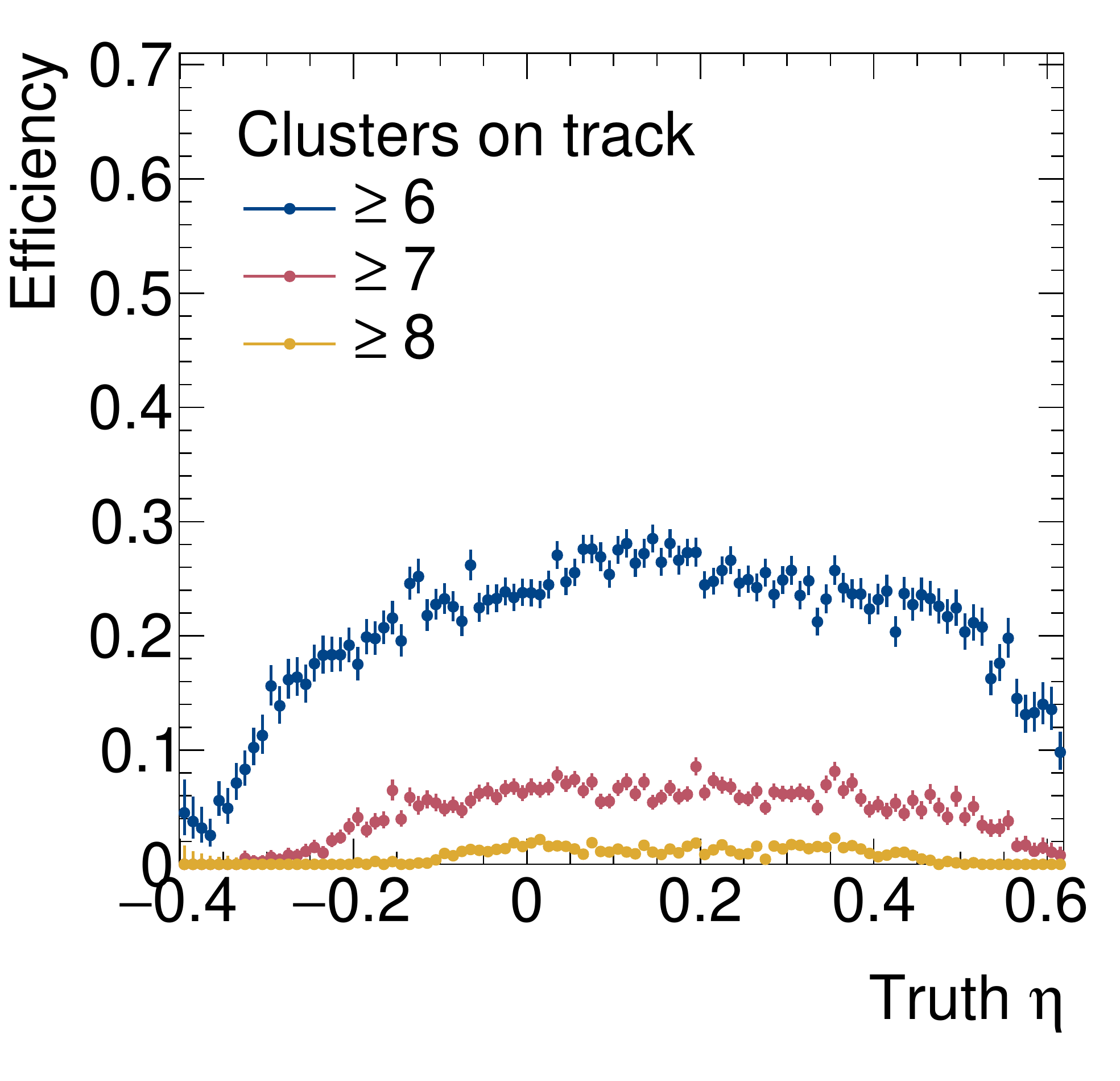}
  \end{minipage}
  \hspace{1em}
  \begin{minipage}{0.4\linewidth}
  \includegraphics[width=\columnwidth]{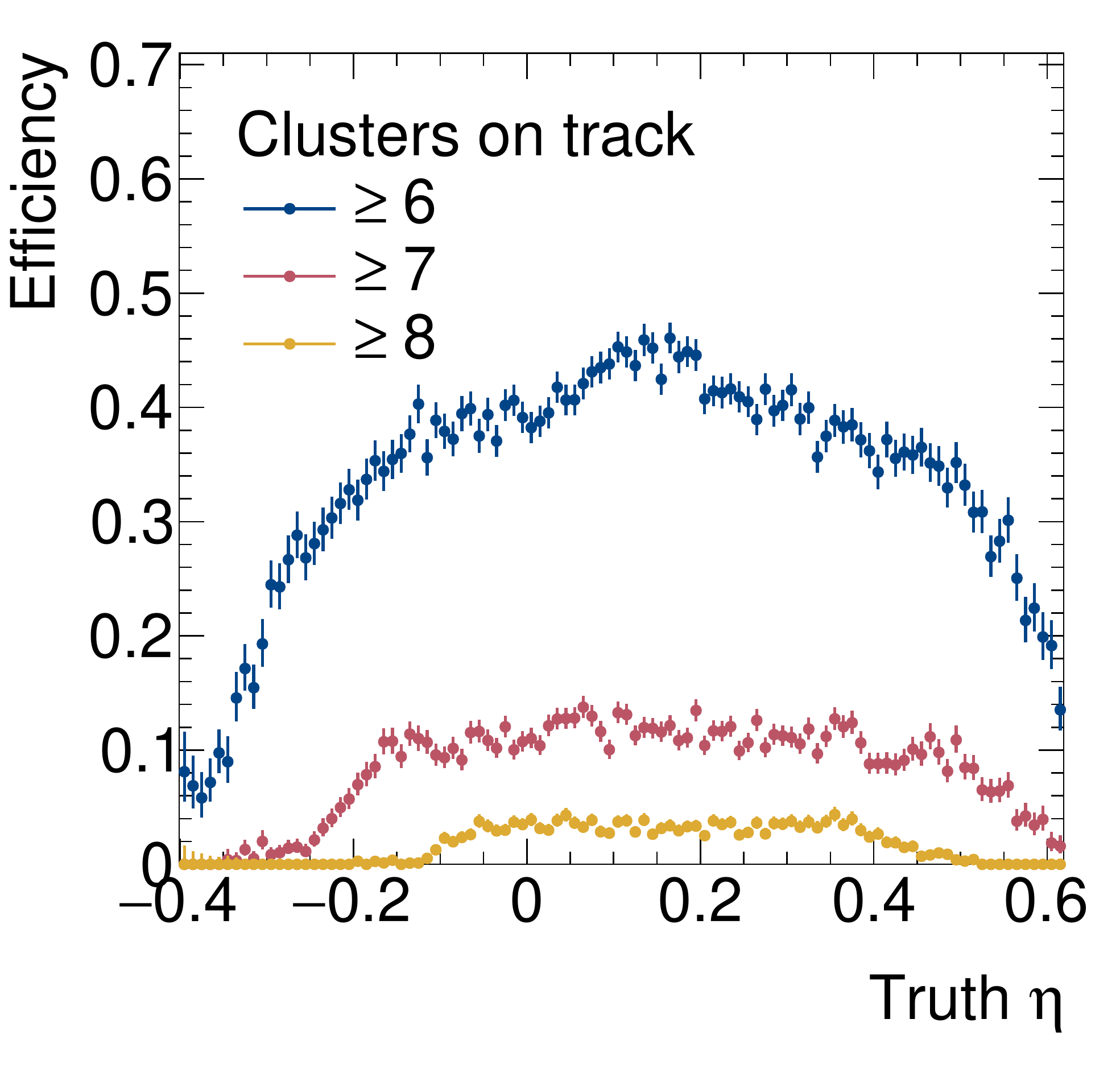}
  \end{minipage}
  \caption{Efficiency of finding at least 6, 7, or 8 hits in unique layers out of eight, as a function of $\eta$, for configuration 1 (left) and 2 (right), Hough transform method.}
  \label{fig:eff_vs_eta}
\end{figure}

\begin{figure}[!ht]
  \centering
  \begin{minipage}{0.4\linewidth}
  \includegraphics[width=\columnwidth]{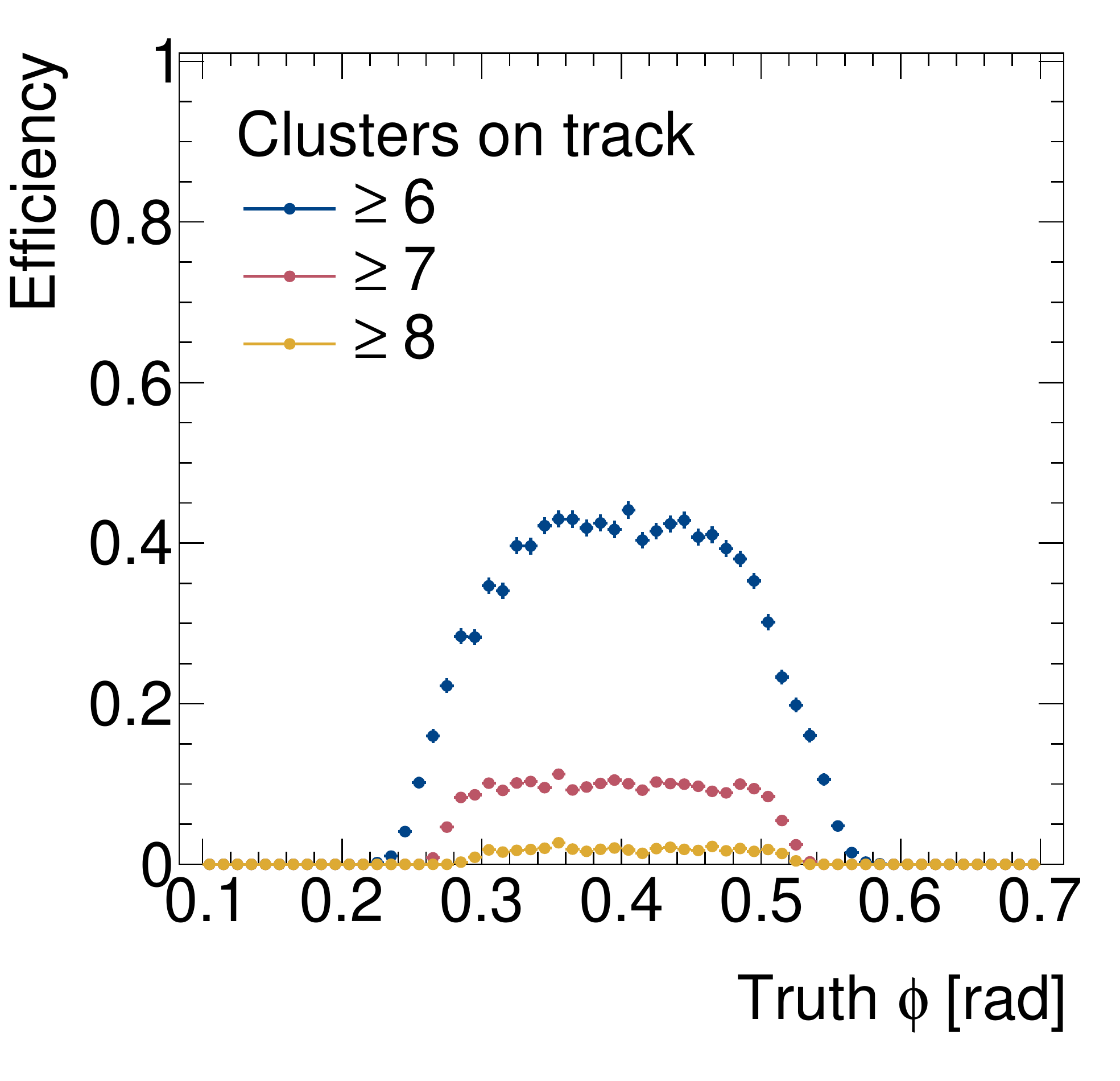}
  \end{minipage}
  \hspace{1em}
  \begin{minipage}{0.4\linewidth}
  \includegraphics[width=\columnwidth]{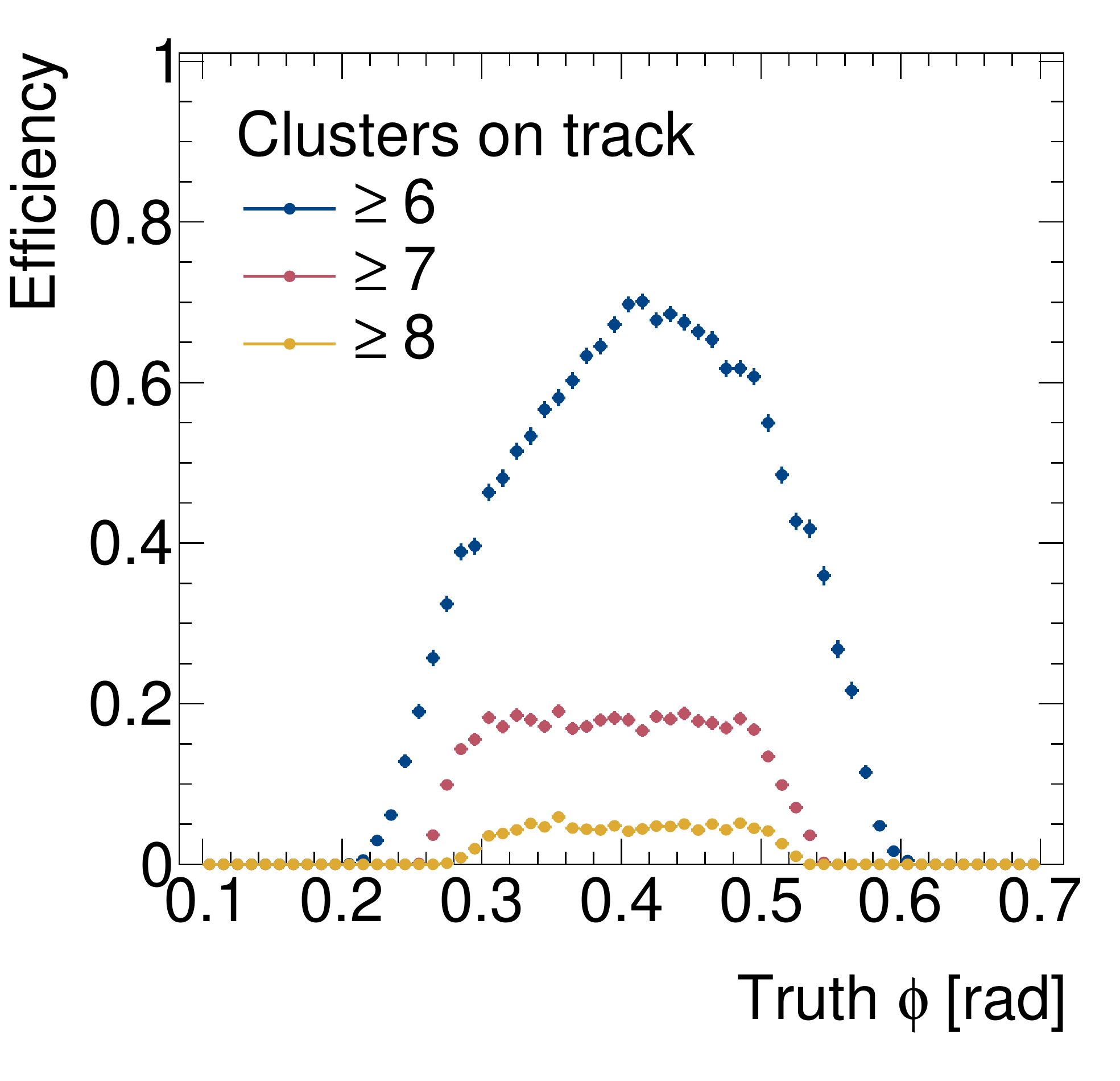}
  \end{minipage}
  \caption{Efficiency of finding at least 6, 7, or 8 hits in unique layers out of eight, as a function of $\phi_0$, for configuration 1 (left) and 2 (right), Hough transform method.}
  \label{fig:eff_vs_phi}
\end{figure}

\begin{figure}[!ht]
  \centering
  \begin{minipage}{0.4\linewidth}
    \includegraphics[width=\columnwidth]{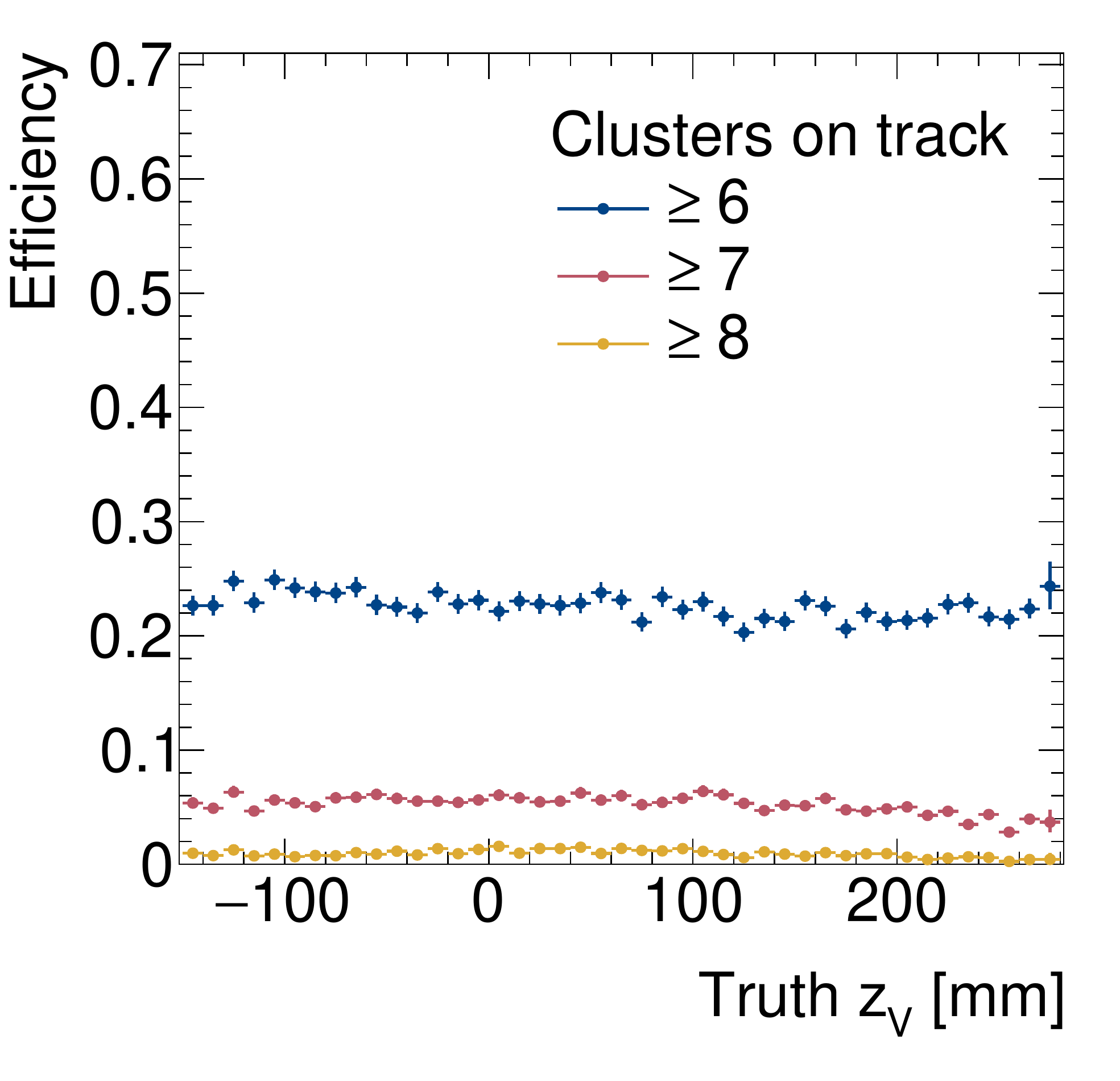}
  \end{minipage}
  \hspace{1em}
  \begin{minipage}{0.4\linewidth}
  \includegraphics[width=\columnwidth]{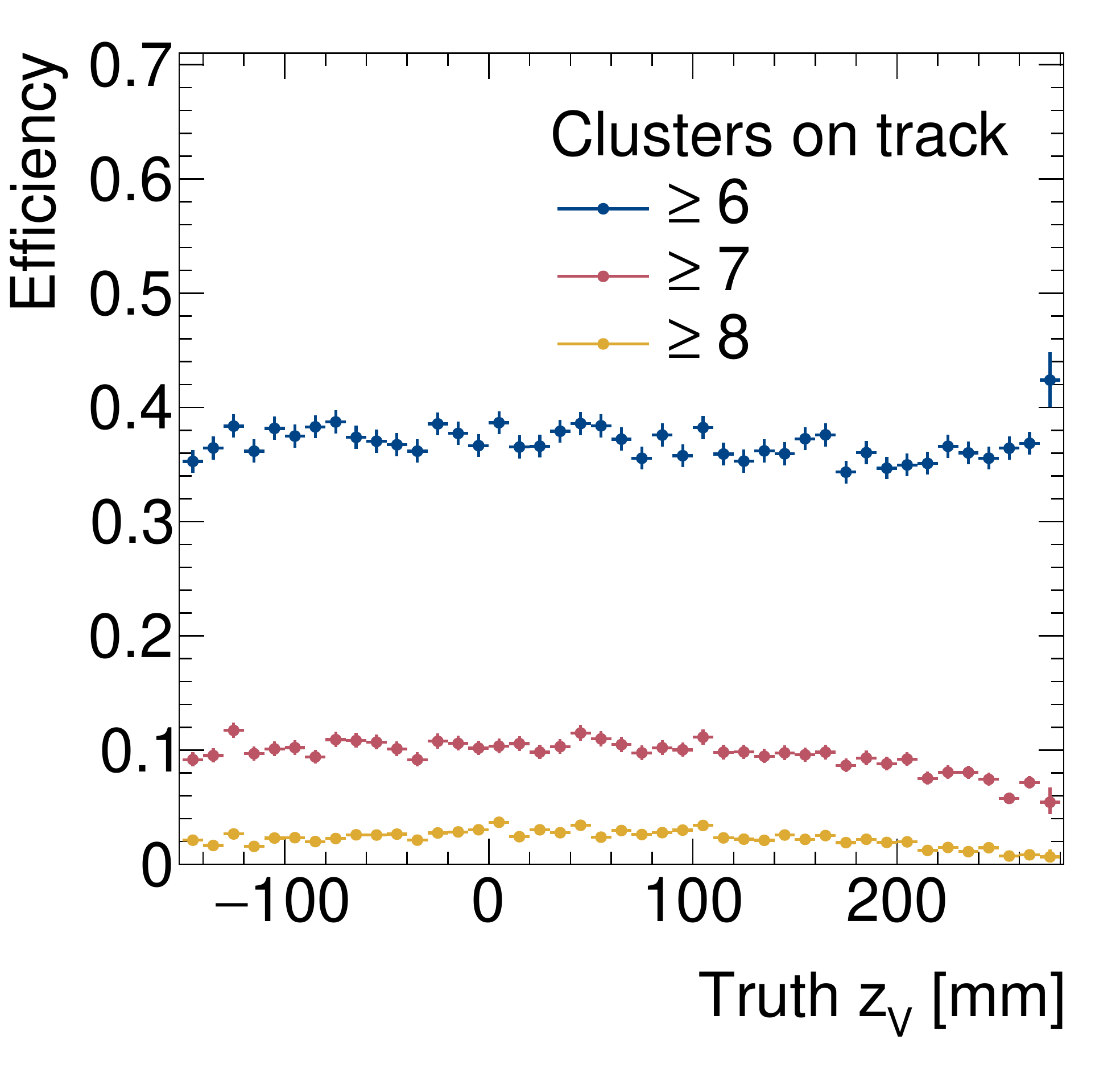}
  \end{minipage}
  \caption{Efficiency of finding at least 6, 7, or 8 hits in unique layers out of eight, as a function of $\phi_0$, for configuration 1 (left) and 2 (right), Hough transform method.}
  \label{fig:eff_vs_z0}
\end{figure}

In the main body of the paper, the efficiency distributions are only presented as a function of $p_\text{T}$ and transverse vertex displacement. Here, the distributions for $\eta_0$, $\phi_0$ and $z_0$ are presented in figures \ref{fig:eff_vs_eta}, \ref{fig:eff_vs_phi}, and \ref{fig:eff_vs_z0} respectively.

\clearpage
\section{Additional efficiency distributions for the pattern matching method}
\label{app:pattern_additional_distribution}

\begin{figure}[!ht]
  \centering
  \begin{minipage}{0.4\linewidth}
  \includegraphics[width=\columnwidth]{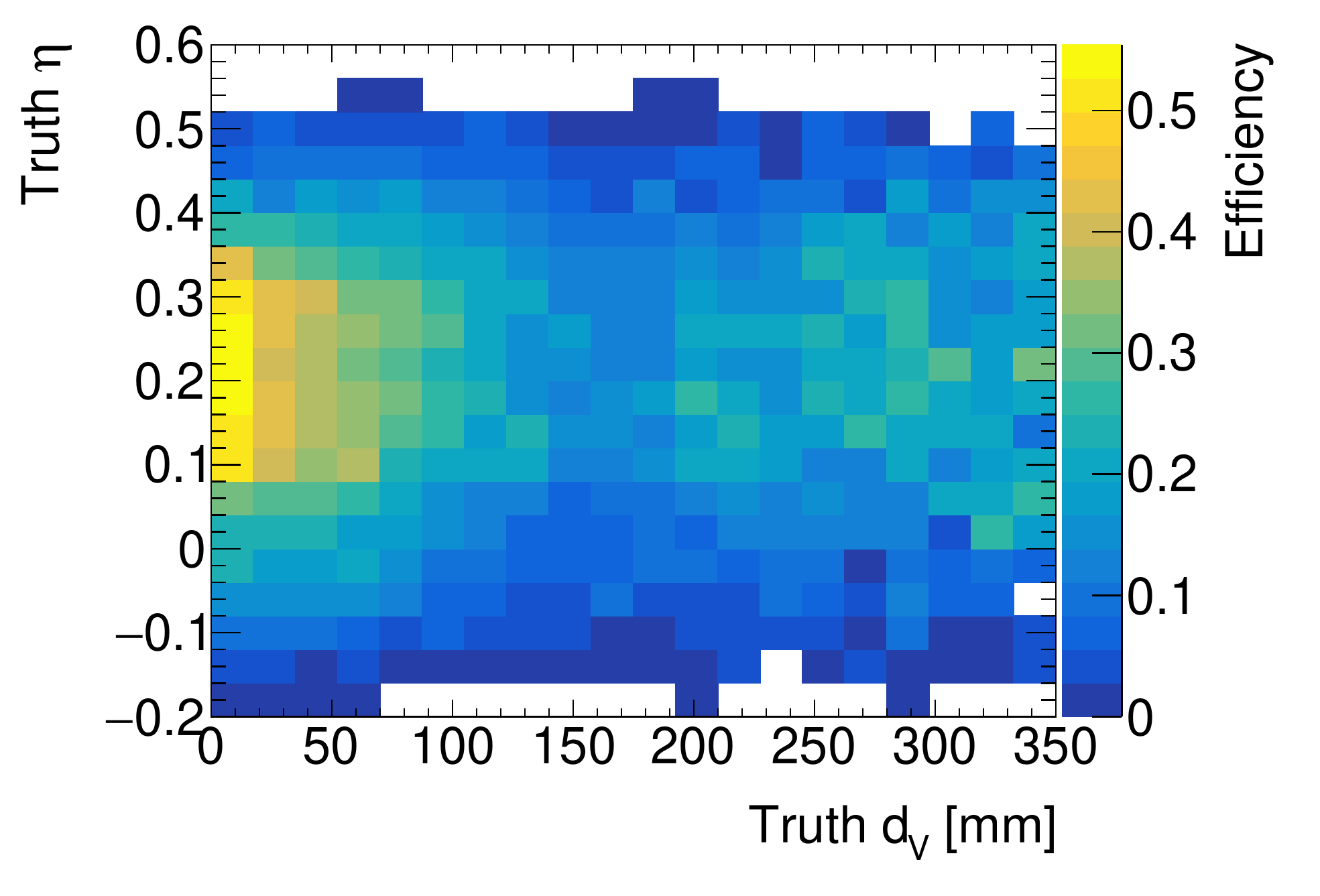}
  \end{minipage}
  \hspace{1em}
  \begin{minipage}{0.4\linewidth}
  \includegraphics[width=\columnwidth]{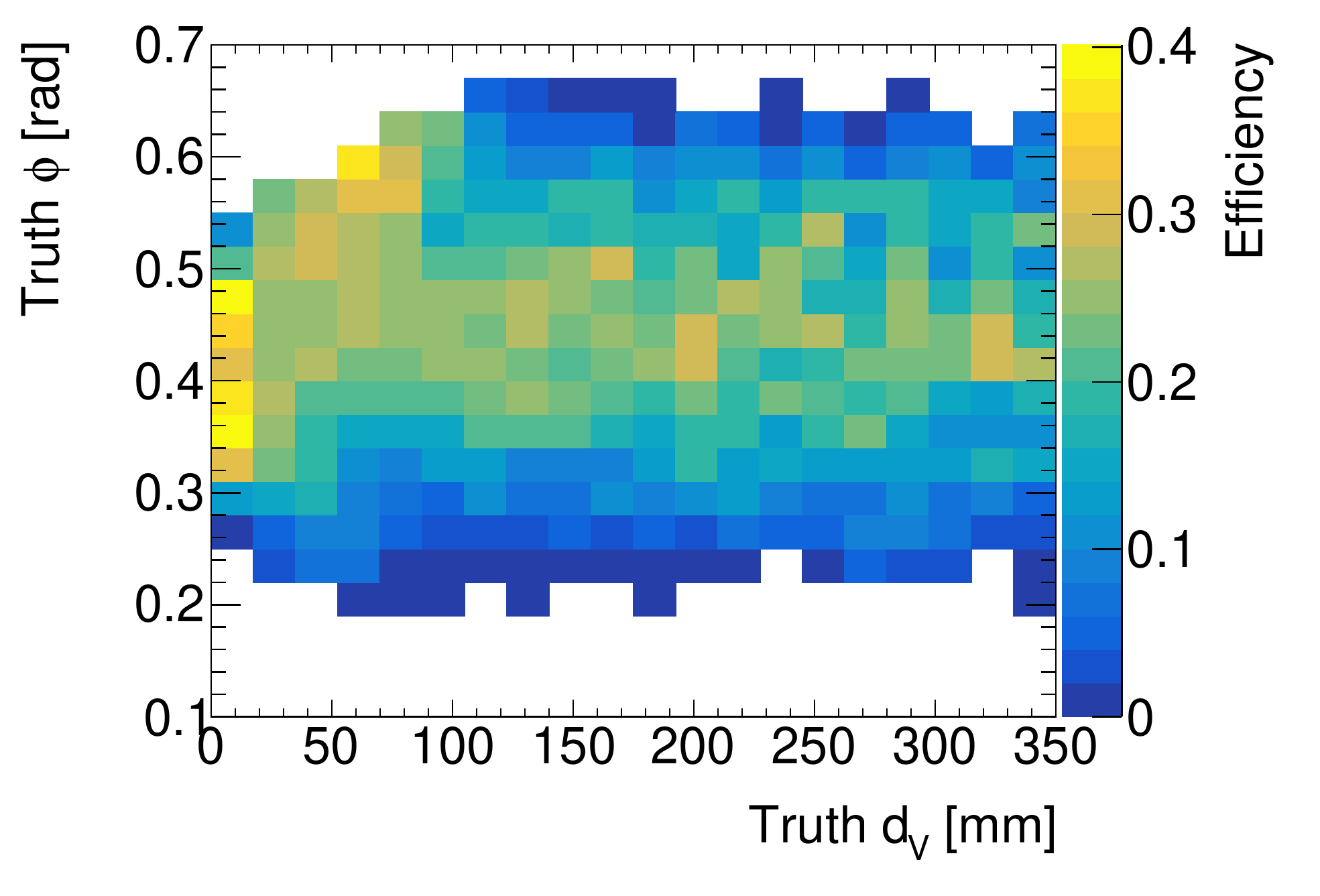}
  \end{minipage}
  \caption{Efficiency of matching at least 6 hits in unique layers out of 8, as a
  function of $d_V$ and $\eta$ (Left), and $d_V$ and $\phi$ (Right), using 1M patterns
  trained on prompt tracks and a SSW of 32, pattern matching method.}
\end{figure}

\begin{figure}[!ht]
  \centering
  \begin{minipage}{0.4\linewidth}
  \includegraphics[width=\columnwidth]{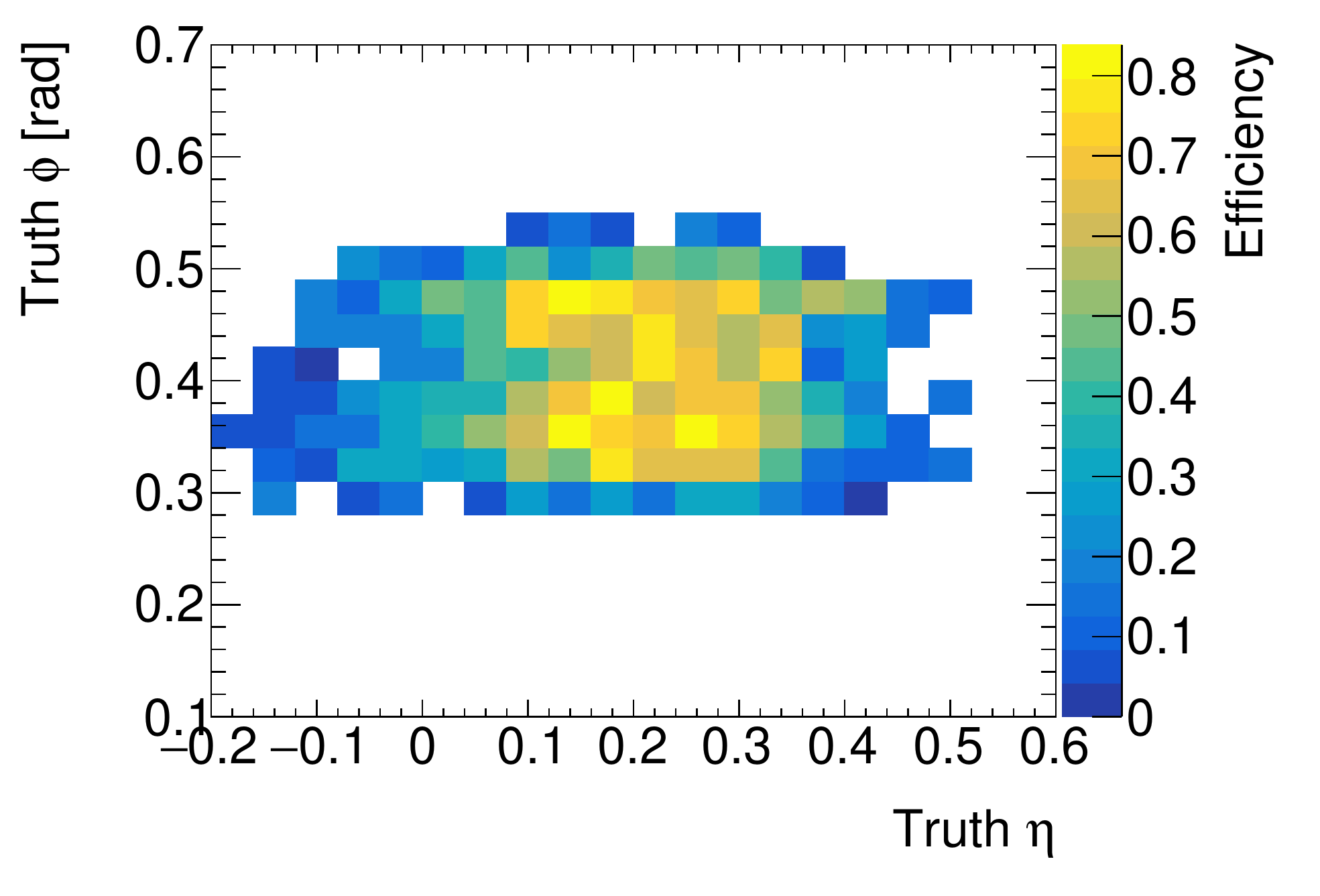}
  \end{minipage}
  \hspace{1em}
  \begin{minipage}{0.4\linewidth}
  \includegraphics[width=\columnwidth]{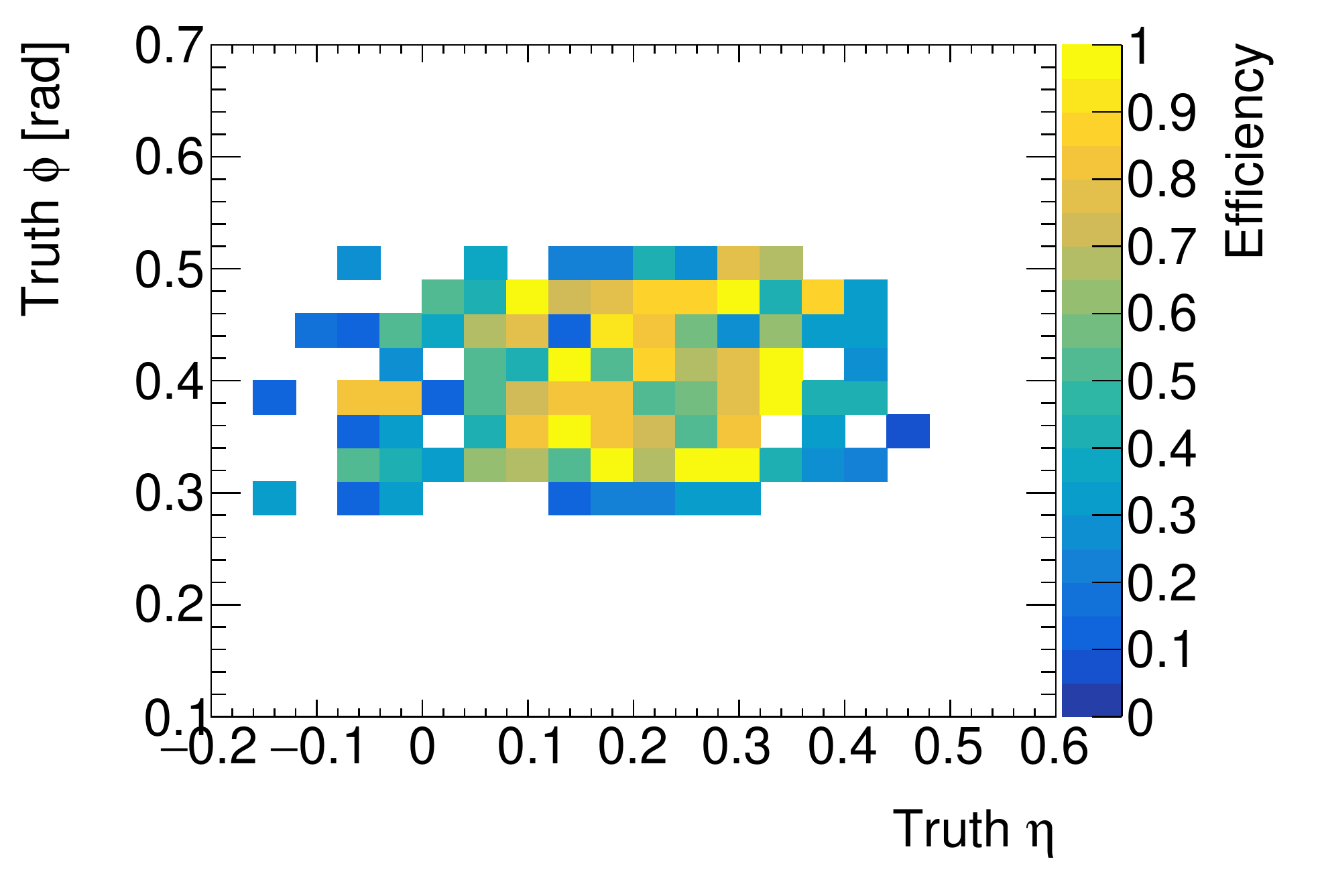}
  \end{minipage}
  \caption{Efficiency of matching at least 6 hits in unique layers out of 8, as a
      function of $\eta$ and $\phi$, requiring $d_V < \SI{10}{\milli\meter}$ (Left),
      and $d_V < \SI{2}{\milli\meter}$ (Right), using 1M patterns
        trained on prompt tracks and a SSW of 32, pattern matching method.}
\end{figure}

\begin{figure}[!ht]
  \centering
  \includegraphics[width=0.6\columnwidth]{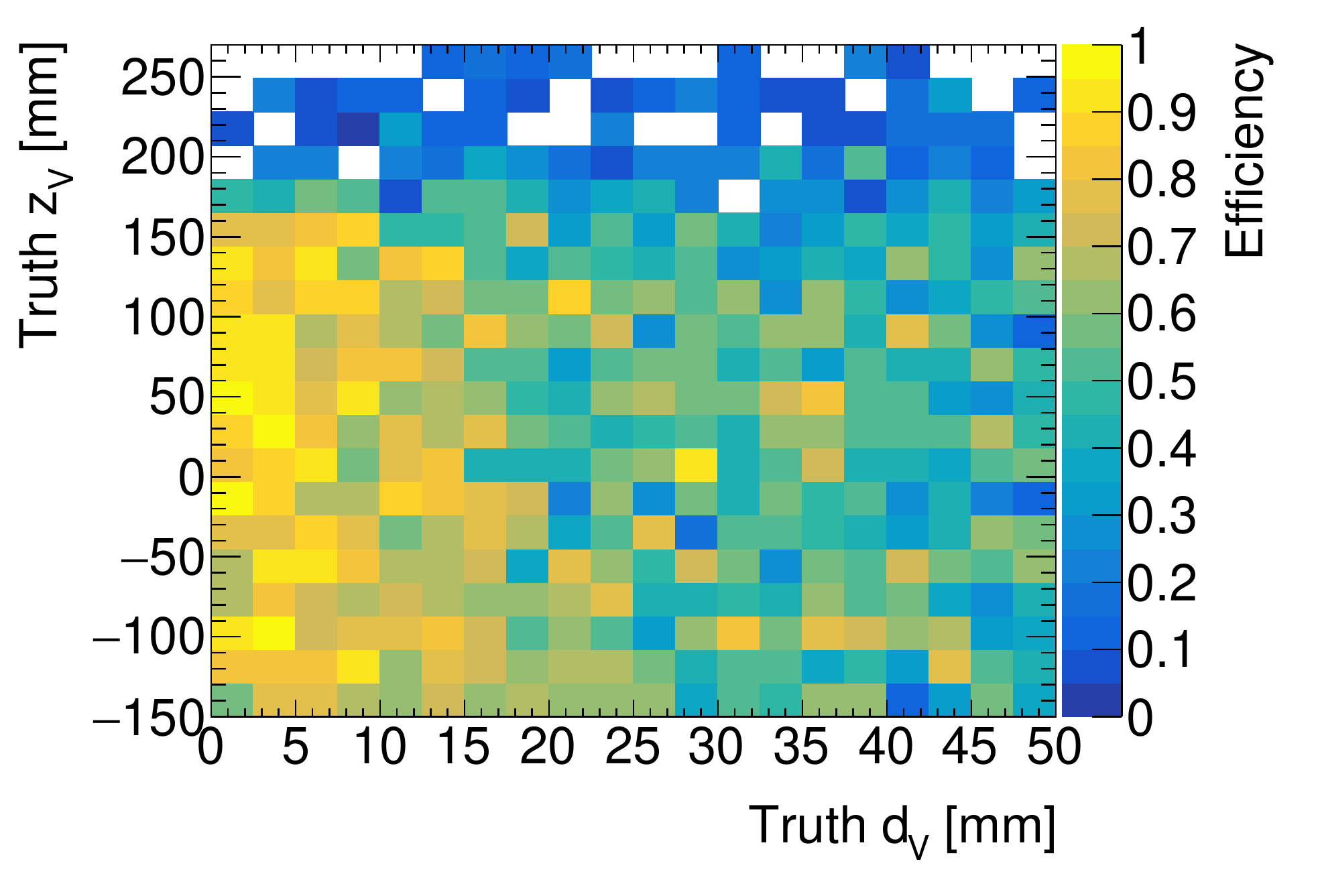}
  \caption{Efficiency of matching at least 6 hits in unique layers out of 8, as
  a function of $d_V$ and $z_V$, requiring $0.1 < \eta < 0.3$ and $0.3 < \phi <
  0.5$, using 1M patterns trained on prompt tracks and a SSW of 32,
  pattern matching method.}
\end{figure}
\end{document}